\documentstyle[11pt,aaspp4,eqsecnum,epsf]{article}

\begin{document}

\title{GRAVITATIONAL COLLAPSE AND STAR FORMATION IN LOGOTROPIC AND
NON-ISOTHERMAL SPHERES}
\author{Dean E. McLaughlin and Ralph E. Pudritz}
\affil{Department of Physics and Astronomy, McMaster University \\
Hamilton, Ontario L8S 4M1, Canada \\ dean@physics.mcmaster.ca,
pudritz@physics.mcmaster.ca}
\lefthead{McLaughlin \& Pudritz}
\righthead{Logotropic Collapse and Star Formation}

\begin{abstract}

We present semi-analytical similarity solutions for the inside-out,
expansion-wave collapse of initially virialized gas clouds with
non-isothermal equations of state.  Results are given for the family of
negative-index polytropes ($P\propto\rho^{\gamma}$, $\gamma\le1$), but we
focus especially on the so-called logotrope, $P/P_c=1+A\ {\hbox{ln}}(\rho/
\rho_c)$. In a separate paper, we have shown this to be the best available
phenomenological description of the internal structure and average properties
of molecular clouds and dense clumps of both high and low mass. The formalism
and interpretation of the present theory are extensions of those in Shu's
(1977) standard model for accretion in self-gravitating isothermal spheres: a
collapse front moves outwards into a cloud at rest, and the gas behind it
falls back to a collapsed core, or protostar. The infalling material
eventually enters free-fall, so that the density profiles and velocity fields
have the same shape ($\rho\propto r^{-3/2}$ and $-v\propto r^{-1/2}$) at small
radii in logotropic and isothermal spheres both. However, several differences
arise from the introduction of a new equation of state. The accretion rate
onto a protostar is not constant in a logotrope, but grows as $\dot{M}\propto
t^3$ during the expansion wave. Thus, the formation time for a star of mass
$M$ scales as $M^{1/4}$; low-mass stars are accreted over longer times, and
high-mass stars over shorter times, than expected in isothermal clouds. This
result has implications for the form and origin of the stellar IMF. In
addition, the gas density behind an expansion wave increases with time in our
theory, but would decrease in an isothermal sphere. The infall velocities also
grow, but at an initially much slower rate than found in an isothermal
collapse. These results apply to low- and high-mass star formation alike. We
briefly discuss how they lead to older inferred collapse ages for Class 0
protostars in general, and for the Bok globule B335 in particular.

\

{\large{\it To appear in The Astrophysical Journal}}

\end{abstract}


\section{Introduction}

One of the cornerstones in any theory of star formation has to be a correct
description of the gravitational collapse of gaseous clouds. The pioneering
studies of \markcite{bod68}Bodenheimer \& Sweigart (1968),
\markcite{lar69}Larson (1969), and \markcite{pen69}Penston (1969) gave
numerical integrations of the isothermal collapse of uniform-density spheres,
and \markcite{lar69}Larson and \markcite{pen69}Penston both went so far as to
derive semi-analytical similarity solutions (descriptions of a fluid flow in
which the densities and velocities approach invariant forms) for the evolution
of their spheres. In a seminal paper which defined the current paradigm for
``inside-out'' collapse and low-mass star formation, \markcite{shu77}Shu
(1977) derived a different, one-parameter family of similarity solutions for
the isothermal problem. \markcite{hun77}Hunter (1977) found still more, and
\markcite{whi85}Whitworth \& Summers (1985) finally pointed out that there is
in fact a two-dimensional continuum of such solutions. Common to all of these
is the eventual formation of a central point mass which subsequently grows as
the cloud continues to fall onto it. This {\it core} is naturally identified
as a protostar (or star and disk).

One of \markcite{shu77}Shu's (1977) solutions in particular --- the so-called
expansion wave --- has played a central role in the development of the
standard theory of star formation, as reviewed, for example, by
\markcite{shu87}Shu, Adams, \& Lizano (1987). In part, this is because of the
simple and intuitive nature of the expansion wave, and because it lends itself
well to the (perturbative) inclusion of rotation (\markcite{ter84}Tereby, Shu,
\& Cassen 1984) and mean magnetic fields (Galli \& Shu \markcite{ga93a}1993a,
\markcite{ga93b}b) in isothermal clouds. In addition, however, it is
physically relevant because it describes the relatively gentle collapse, from
the center on out, of a sphere which is initially in virial equilibrium. As
we discuss shortly, the idea that pre-collapse interstellar clouds are in
equilibrium configurations is one which meshes with a variety of observations.
There are contrary arguments (e.g., \markcite{bon96}Bonnell, Bate, \& Price
1996) that star-forming clouds should collapse from non-equilibrium, or even
near-uniform conditions; but this is a violent process (the flow of
\markcite{lar69}Larson [1969] and
\markcite{pen69}Penston [1969] is everywhere highly supersonic) which would
give rise to strong kinematic signatures that have never been observed.
Rather, the most convincing claim to a direct detection of collapse has been
made for the Bok globule B335, where molecular line profiles are consistent
with the more moderate velocity field expected for an expansion wave
(\markcite{zho92}Zhou 1992; \markcite{zho93}Zhou et al.~1993).

The standard theory has its difficulties, however. For instance,
\markcite{shu77}Shu's (1977) model describes the collapse of highly unstable,
{\it singular} isothermal spheres, and is not necessarily applicable to more
realistic clouds with finite central densities (such as the marginally stable
Bonnor-Ebert sphere: \markcite{hun77}Hunter 1977; \markcite{fos93}Foster \&
Chevalier 1993). As another example, an isothermal expansion wave leads to a
time-invariant protostellar accretion rate, $\dot{M}=0.975\sigma^3/G$, which
appears to be inconsistent with the observed luminosities of young stellar
objects in the Taurus-Auriga star-forming region specifically (e.g.,
\markcite{ken94}Kenyon et al.~1994). But most important is the fact that an
isothermal equation of state, while a natural and instructive first
approximation, is ultimately incompatible with observations of both giant
molecular clouds (GMCs) and the subcondensations, the dense clumps where stars
are ultimately born, within them.

As suggested above, GMC complexes, and individual clumps, are well described
by models of (magnetic) virial-equilibrium spheres which are ``truncated'' by
a constant surface pressure exerted by a diffuse external medium (e.g.,
\markcite{mye88}Myers \& Goodman 1988; \markcite{ber92}Bertoldi
\& McKee 1992; \markcite{elm89}Elmegreen 1989; \markcite{mcl96}McLaughlin \&
Pudritz 1996). (The dense clumps are often referred to as cores, but here we
reserve that term for the central point mass, or protostar, which develops in
a collapsing sphere.) The virial balance relies on superthermal random
velocities which {\it increase outwards} within any given clump. The
nonthermal component $\sigma_{\rm{NT}}$ of the total velocity dispersion
(linewidth) is larger in higher-mass clumps (\markcite{cas95}Caselli \& Myers
1995), but is present at even the smallest observed radii in low-mass clumps
as well (\markcite{ful92}Fuller \& Myers 1992). Indeed, $\sigma_{\rm{NT}}$
grows more rapidly with increasing radius in low-mass clumps than in high-mass
ones, and {\it no interstellar cloud obeys an isothermal equation of state}.

The first attempts to model the nonthermal linewidths (e.g.,
\markcite{mye92}Myers \& Fuller 1992; \markcite{liz89}Lizano \& Shu 1989)
tended to focus on the fact that $\sigma_{\rm{NT}}$ is small in low-mass
clumps. \markcite{mye92}Myers \& Fuller, in particular, construct equilibrium,
self-gravitating spheres from a gas equation of state (EOS) which is
explicitly isothermal at small radii in a cloud, and becomes progressively
``softer'' ($P$ depends less than linearly on $\rho$, so that
$\sigma^2=P/\rho$ increases with decreasing density) further out. Because the
clumps in GMCs can be viewed as pressure-truncated spheroids, with truncation
at a smaller radius corresponding to a lower mass, this approach effectively
ensures that the low-mass clumps are still treated as (very nearly) isothermal
spheres. Any consequences which the nonthermal support might have for collapse
and star formation are then largely confined to the high-mass regime. For
example, \markcite{mye92}Myers \& Fuller (1992) estimate the formation times
of stars in their model, and conclude that high-mass stars are accreted
relatively faster than they would be under an isothermal EOS, while low-mass
stars grow at roughly the same rate.

By contrast, \markcite{mcl96}McLaughlin \& Pudritz (1996=MP96) have adopted a
phenomenological, {\it logotropic} EOS --- $P/P_c=1+A\ {\hbox{ln}}(\rho/
\rho_c)$ --- for molecular clouds and clumps. This EOS provides a unified
treatment of low- and high-mass clumps and entire GMCs. It can account
quantitatively for (1) the global size-linewidth and mass-radius relations
(e.g., \markcite{lar81}Larson 1981) {\it between} cloud complexes; (2) the
typical mass and density contrasts between GMCs and the clumps within them;
and (3) the observed dependence of linewidth on radius {\it inside} dense
clumps of any mass. Moreover, the form of the EOS itself is never close, even
on small spatial scales or in low-mass clouds, to the isothermal $P\propto
\rho$. The details of {\it both} low- {\it and} high-mass star formation in a
logotrope will therefore differ from the standard results for isothermal
collapse. (It should be noted that our logotrope is different from that
introduced by \markcite{liz89}Lizano \& Shu [1989], which essentially has
$P\sim \rho + {\hbox{ln}}\rho$, and cannot account for all three of the
properties of interstellar clouds listed here.)

In this paper, then, we derive semi-analytical similarity solutions for the
collapse of nonmagnetic, non-rotating gas spheres with softer-than-isothermal
equations of state. We concentrate on the case of a logotropic EOS as the most
self-consistent description of the ubiquitous turbulence in interstellar
clouds; but in a pair of Appendices, results are also given for the more
general family of negative-index polytropes ($P\propto\rho^{\gamma}$, with
$0<\gamma\le 1$). (Although the self-similar collapse of polytropes has
previously been investigated by, e.g., \markcite{sut88}Suto \& Silk [1988], we
also briefly discuss the accretion rates and star formation timescales in such
spheres.)

Section 2 begins with a brief summary of the equilibrium structure of a
logotrope (\markcite{mcl96}MP96), and then goes on to an analysis of its
collapse. Our development owes much to the physical discussion of
\markcite{shu77}Shu (1977): we first find a class of collapse solutions for
singular, $\rho\propto r^{-1}$ spheres which are very accurate approximations
to more realistic, non-singular logotropes, and then focus on the inside-out,
expansion-wave collapse of an initially virialized cloud. Two main features
are a post-collapse density profile and velocity field which have the same
free-fall shape ($\rho\propto r^{-3/2}$ and $-u\propto r^{-1/2}$ at small
radii) found for an isothermal expansion wave, {\it but} a rate of accretion
onto the collapsed core which is no longer constant in time; rather,
$\dot{M}\propto t^3$ here.  These results allow for an evaluation, in \S3, of
the formation times for stars of various masses.  High-mass stars will be
accreted more rapidly in a logotrope than in an isothermal sphere, but a
low-mass star in a logotrope will appear later than a star of the same mass
would in an isothermal sphere.  This ``squeezing'' of formation timescales,
which we estimate to be typically of order $1-3\times10^6$ yr, should have
important implications for the origin and shape of the stellar IMF. In \S4 we
consider the collapse of a 1 $M_\odot$ logotrope specifically, and briefly
discuss relevant observations of two young stellar objects, NGC 2071 and B335.
We conclude that the age of B335 is probably greater than $10^6$ yr, an order
of magnitude older than suggested by the application of isothermal collapse
models. Since B335 is one of the family of the youngest known, ``Class 0''
protostars (\markcite{and93}Andr\'e, Ward-Thompson, \& Barsony 1993), these
objects might generally be rather older than is currently believed.

\section{Similarity Solutions}

\subsection{The Logotrope}

Our logotropic EOS (\markcite{mcl96}MP96) is
\begin{equation}
{P\over{P_c}}=1 + A\ {\hbox{ln}}\left({\rho\over{\rho_c}}\right)\ ,
\label{eq:24}
\end{equation}
where $P_c$ and $\rho_c$ are the central pressure and density in a cloud, and
$A$ is essentially a free parameter. The total velocity dispersion
$\sigma^2=P/\rho$ increases outwards from $r=0$ until it reaches a maximum at
some large radius, beyond which it decreases somewhat. Since linewidths are
essentially purely thermal on the smallest scales in real interstellar clouds,
and are also observed to increase with radius (e.g., \markcite{ful92}Fuller \&
Myers 1992; \markcite{cas95}Caselli \& Myers 1995), the central velocity
dispersion in a logotrope is identified with the thermal value:
$\sigma_c^2\equiv P_c/\rho_c=kT/\mu m_H$. Once a value of $A$ is specified,
the EOS (\ref{eq:24}) can be used to integrate a Lane-Emden type of equation
for a self-gravitating, virialized gas sphere. This then allows for a
quantitative comparison of the {\it internal} $\sigma$ vs.~$r$ profiles of
real and model clouds. Good agreement with data on GMC clumps, of low and high
mass alike, is obtained for $A\simeq0.2\pm0.02$.

These points are discussed in more detail in \markcite{mcl96}MP96. Starting
from the premise that interstellar clouds are pressure-truncated --- their
boundaries defined by a surface pressure $P_s$ due to a more diffuse,
surrounding medium --- that paper develops expressions for the masses and
densities of virial-equilibrium gas spheres with arbitrary equations of state.
Quite general arguments then imply that GMCs, and the {\it largest} clumps
within them, are all at or near their critical masses: if they were any more
massive, given their temperatures, surface pressures, and magnetic field
strengths, they would be unstable to radial perturbations (see also
\markcite{mck89}McKee 1989; \markcite{ber92}Bertoldi \& McKee 1992). Although
this conclusion is independent of the exact form of the gas EOS, not all
equations of state provide for such a critical mass. For example, the family
of negative-index polytropes (in which $P\propto\rho^{\gamma}$, with
$\gamma<1$) also have velocity dispersions that increase with radius, but are
{\it always stable} against radial perturbations in the analysis of
\markcite{mcl96}MP96.  Put another way, and independently of any stability
analysis, such polytropes cannot simultaneously account for the internal
structure of molecular clumps and the global properties
(\markcite{lar81}Larson's [1981] laws: $M\propto R^2$ and
$\sigma_{\rm{ave}}\propto R^{1/2}$) of GMCs; the logotrope is the simplest EOS
for which this is possible.

The critical mass of an $A=0.2$ logotrope is given by equation (2.11) of
\markcite{mcl96}MP96:
\begin{equation}
M_{\rm{crit}}={{250}\over{\alpha^{3/2}}}\ M_{\odot}\ \left({T\over{10\
{\hbox{K}}}}\right)^2 \left({{P_s}\over{1.3\times10^5\ k\ {\hbox{cm}}^{-3}\ 
{\hbox{K}}}}\right)^{-1/2}\ ,
\label{eq:39}
\end{equation}
where we have explicitly used the fact that $\sigma_c^2=kT/\mu m_H$, and set
the mean molecular weight to $\mu=2.33$.  The constant $\alpha$ here depends
on the strength of any mean magnetic field threading the cloud.  If there is
equipartition between kinetic and mean magnetic energies, then
$\alpha\approx1$; in the absence of any mean field at all, $\alpha=35/18$ for
the logotrope and $M_{\rm{crit}}\simeq92M_\odot$ for this temperature and
surface pressure. A smaller value of $A$ results in a larger critical mass:
for example, $A=0.18$ would raise $M_{\rm{crit}}$ by a factor of about 2.5.
\markcite{mcl96}MP96 also derive the mean density of a critical-mass
logotrope:
\begin{equation}
\rho_{\rm{ave,crit}}=1.31\times10^{-20}\ {\hbox{g cm}}^{-3}\ 
\left({T\over{10\ {\hbox{K}}}}\right)^{-1}
\left({{P_s}\over{1.3\times10^5\ k\ {\hbox{cm}}^{-3}\ {\hbox{K}}}}\right)\ .
\label{eq:40}
\end{equation}

The fiducial $P_s$ chosen here is appropriate for dense clumps embedded in
self-gravitating GMCs, which amplify the pressure of the hot ISM by about an
order of magnitude (\markcite{elm89}Elmegreen 1989; \markcite{ber92}Bertoldi
\& McKee 1992); it is also similar to the value adopted by \markcite{shu77}Shu
(1977) in his study of isothermal collapse. It is significant that the
logotropic $M_{\rm{crit}}$ is vastly greater than the $\sim$$1M_\odot$ mass of
a critical, $T=10$ K isothermal sphere under the same surface pressure.  This
is part of what makes our model more consistent with observations of the ISM,
and in particular with the arguments, mentioned above, that only the largest
clumps in GMCs are at their critical mass, with smaller ones being in a stable
virial equilibrium.

Because we are dealing with pressure-truncated spheres, a cloud's mass can be
viewed as being set by the dimensionless ``truncation radius'' $R/r_0$ (where
$r_0\equiv3\sigma_c/[4\pi G\rho_c]^{1/2}$) at which the internal pressure
equals that of an ambient medium. That is, there is an EOS-dependent $R/r_0$
which corresponds to a critical-mass cloud; any smaller $R/r_0$ is equivalent
to a unique $M/M_{\rm{crit}}<1$. Indeed, $R/r_0$, along with a surface
pressure $P_s$ and kinetic temperature $T$, determines all the properties of a
cloud (see Appendix A of \markcite{mcl96}MP96). An example of this is given in
Table \ref{tab0}, where we list the center-to-edge density and pressure
contrasts, mean density, and mass-averaged linewidth (thermal plus nonthermal:
$\sigma_{\rm{ave}}^2= P_{\rm{ave}}/\rho_{\rm{ave}}$) for logotropes with
$A=0.2$. The absolute values of the masses and free-fall times there
($\overline{t}_{\rm{ff}}= [3\pi/32G\rho_{\rm{ave}}]^{1/2}$) apply to purely
hydrostatic spheres with no mean magnetic fields, as these are the types of
objects which our collapse calculations will directly address. If $P_s$ or $T$
is changed, the mass and mean density of the critical cloud will be affected
as in equations (\ref{eq:39}) and (\ref{eq:40}). Then, because the
dimensionless $R/r_0$ actually determines the mass and free-fall time relative
to the critical values, these quantities will both change at every radius in
Table \ref{tab0}, and by the same factors as in the critical cloud. In any
case, it is significant that the total linewidth $\sigma_{\rm{ave}}$ is
generally larger in bigger clouds. Lower-mass objects are supported to a
greater, {\it but not complete} extent by thermal motions.

\placetable{tab0}

Figure \ref{fig1} shows the density profile of a critical-mass logotrope, and
compares it with that of a critical-mass, or Bonnor-Ebert, isothermal sphere
(\markcite{bon56}Bonnor 1956; \markcite{ebe55}Ebert 1955;
\markcite{mcc57}McCrea 1957). The structure of less massive spheres can also
be derived from this diagram; for example, the $\rho$ vs.~$r$ curve for a
logotrope with $M/M_{\rm{crit}}\simeq0.33$ is obtained by vertically shifting
the curve in Fig.~\ref{fig1} such that $\rho/\rho_s=1$ at $r/r_0\simeq9.4$
(cf.~Table \ref{tab0}). Also shown in the Figure, as broken lines, are the
analytical singular solution to the equation of hydrostatic equilibrium for a
self-gravitating logotrope (MP96),
\begin{equation}
\rho(r)=(AP_c/2\pi G)^{1/2}\ r^{-1}\ ,
\label{eq:20}
\end{equation}
and the singular solution $\rho(r)=(\sigma_c^2/2\pi G)\ r^{-2}$ for an
isothermal sphere. Equation (\ref{eq:20}) closely approximates the bounded
logotrope (finite central density; solid line) at all $r/r_0\ga0.4$, and thus
adequately describes the structure even of stable low-mass spheres.  On the
other hand, the singular isothermal sphere bears very little relation to the
Bonnor-Ebert configuration. (The logotropic prediction $\rho\propto r^{-1}$ is
one with some observational support. This is discussed in
\markcite{mcl96}MP96; here we note only that an observed column density
profile $N\propto r^{-1}$, which is not uncommon, does {\it not} necessarily
imply $\rho\propto r^{-2}$.)

\placefigure{fig1}

Since a low-mass clump (say, $M_{\rm{tot}}\sim1M_\odot$) is expected to be
globally stable against radial perturbations, how does it come to collapse
and form a star?  When magnetic fields are present, ambipolar diffusion should
induce a slow contraction of the clump, during which its initial (time
$t=-\infty$) central concentration will be enhanced and its density profile
brought into agreement with the singular solution (\ref{eq:20}) over
essentially all radii --- a configuration that is manifestly unstable (e.g.,
\markcite{mcl96}MP96). Alternatively, in a purely hydrodynamic situation,
collapse could be initiated by the growth of a local density fluctuation; or
the clump could be put out of equilibrium by the accretion of mass from the
surrounding medium. The time at which a cloud becomes singular is labelled
$t=0$, and corresponds to the formation of a central point mass (or core;
physically, a protostar and disk).  Following \markcite{shu77}Shu (1977), it
is the subsequent, $t\ge0$ evolution of nonmagnetic spheres that we now
examine in detail.

\subsection{Fluid Equations}

We begin with the usual Eulerian equations for spherically symmetric flow in
a self-gravitating fluid (cf.~\markcite{lar69}Larson 1969; \markcite{shu77}Shu
1977):
\begin{eqnarray}
{{\partial u}\over{\partial t}} + u {{\partial u}\over{\partial r}} +
{{GM}\over{r^2}} + {{dP}\over{d\rho}}{{\partial\ {\hbox{ln}}\rho}\over
{\partial r}} & = & 0 \label{eq:21} \\
{{\partial M}\over{\partial t}} + 4\pi r^2 \rho u & = & 0 \label{eq:22} \\
{{\partial M}\over{\partial r}} - 4\pi r^2 \rho & = & 0 \ , \label{eq:23}
\end{eqnarray}
where $M(r,t)$ is the {\it total} mass (including any core) inside radius $r$
at time $t$, and $u(r,t)$ is the instantaneous fluid velocity. The gas EOS
influences the collapse through the pressure force $\rho^{-1}\partial P/
\partial r$ in equation (\ref{eq:21}). For the most part, we confine our
attention to the logotropic EOS (eq.~[\ref{eq:24}]), but in Appendices A and B
below we also consider the more general family of negative-index polytropes.

We define a similarity variable
\begin{equation}
x={r\over{a_t t}}\ ,
\label{eq:26}
\end{equation}
where $a_t$ is a function of time with dimensions of velocity. Dimensionless
density, velocity, and mass variables may then be set up as
\begin{equation}
\alpha(x)=4\pi G t^2\rho(r,t)\ \ \ \ \ \ v(x)=u(r,t)/a_t\ \ \ \ \ \ 
m(x)=GM(r,t)/a_t^3 t\ .
\label{eq:27}
\end{equation}
Further, because the square of the sound speed is $dP/d\rho=AP_c/\rho$ for
this EOS, we are led to set
\begin{equation}
a_t=[AP_c(4\pi Gt^2)]^{1/2}\ ,
\label{eq:25}
\end{equation}
where $A\simeq0.2$ for real interstellar clouds, and where the constant $P_c$
refers to the central pressure in the {\it initial}, pre-collapse
($t=-\infty$) cloud (as such, it reflects the constant pressure of the ambient
medium, and the total mass of the cloud; see, e.g., Table \ref{tab0}).
Ultimately, the central core mass grows as $M(0,t)\propto a_t^3t\propto t^4$,
and the accretion rate as $\dot{M}(0,t)\propto t^3$. This is in sharp contrast
to an isothermal collapse, where the accretion rate is constant in time (see
\S3 below for further discussion). In any case, straightforward manipulation
of equations (\ref{eq:21}) -- (\ref{eq:23}) yields
\begin{eqnarray}
\left[(2x-v)^2-{1\over{\alpha}}\right]{{d\ {\hbox{ln}}\alpha}\over{dx}} & = &
\left[{\alpha\over{4}}-{2\over{x}}(2x-v)\right](2x-v) + \left[2(2x-v) + v\right]
\label{eq:29} \\
\left[(2x-v)^2-{1\over{\alpha}}\right]{{dv}\over{dx}} & = &
\left[{\alpha\over{4}}(2x-v)-{2\over{\alpha x}}\right](2x-v) +
\left[{2\over{\alpha}}+v(2x-v)\right]\label{eq:210} \\
m & = & {{\alpha x^2}\over{4}}(2x-v)\ . \label{eq:28}
\end{eqnarray}
These equations are written in a form similar to that of \markcite{shu77}Shu
(1977) for the isothermal sphere, which is included as a special case in the
general development of Appendix A below.

There are two simple analytical solutions of equations (\ref{eq:29}) --
(\ref{eq:28}). The first has a uniform density,
\begin{equation}
\alpha={2\over{3}}\ \ \ \ \ \ v={{2x}\over{3}}\ \ \ \ \ \ 
m={{2 x^3}\over{9}}\ ,
\label{eq:211}
\end{equation}
and is equally valid for any polytropic EOS. Of more interest is the static
(singular) solution,
\begin{equation}
\alpha={{\sqrt{2}}\over{x}}\ \ \ \ \ \ v\equiv0\ \ \ \ \ \ 
m={{x^2}\over{\sqrt{2}}}\ .
\label{eq:212}
\end{equation}
This corresponds to a time-independent density profile, $\rho(r)=(AP_c/2\pi
G)^{1/2}\ r^{-1}$, which is just that found (eq.~[\ref{eq:20}]) to satisfy
the equation of hydrostatic equilibrium for a self-gravitating logotrope.
As discussed in \S2.1, it also describes the large-radius structure of the
bounded sphere --- the type of cloud we expect to begin with at
$t=-\infty$.

An entire class of singular solutions, valid at large radii or small
(positive) times, can be found by taking $\alpha$ and $|v|$ to be much smaller
than $x$ in equations (\ref{eq:29}) -- (\ref{eq:28}). In this $x\rightarrow
\infty$ limit, we have
\begin{equation}
\alpha\rightarrow {C\over{x}}\ \ \ \ \ \ 
v\rightarrow -{C\over{2}}\left(1-{2\over{C^2}}\right)
\ \ \ \ \ \ m\rightarrow {{Cx^2}\over{2}}\ .
\label{eq:213}
\end{equation}
For an inflow scenario ($v<0$), the density normalization $C$ must be greater
than the hydrostatic-equilibrium value $\sqrt{2}$. Each $C>\sqrt{2}$ therefore
corresponds to the collapse of a logotrope which is initially just out of
equilibrium, with densities (and pressures) everywhere enhanced by a factor
$1+\delta=C/\sqrt{2}$. As in the equilibrium case, the cloud starts at
$t=-\infty$ with a finite central density, and $\rho$ at large radii given by
equation (\ref{eq:213}), and then contracts to follow that singular profile
everywhere by $t=0$. These solutions are the analogues of
\markcite{shu77}Shu's (1977) ``minus solutions without critical points'' for
the singular isothermal sphere --- although we show in \S2.3 that in our case,
some of the out-of-equilibrium flows {\it can} pass through critical points.

Although $v$ is a finite constant at $t=0$ here, the dimensional velocity
$u\sim vt$ vanishes, and the cloud is essentially at rest at the time of core
formation. This is consistent with a physical picture at $t<0$ of quasistatic
evolution towards the singular profile. As suggested above, such a slow
contraction could be driven by ambipolar diffusion, and we therefore view the
family of solutions (\ref{eq:213}) as reasonable starting points for the
post-core formation evolution of a collapsing logotrope. This is the same tack
taken by \markcite{shu77}Shu (1977) in his treatment of the isothermal sphere,
but in one important way it is perhaps better justified for our EOS. If it
happens that the mean magnetic field is initially weak and unable to
significantly influence the evolution of a cloud, then the assumption of a
gentle approach to the singular profile, and the validity of equations
(\ref{eq:213}) in the $t=0$ limit of the collapse, have to be questioned

The singular isothermal sphere is a poor description of the critical-mass
Bonnor-Ebert configuration, or of any less massive one (see Fig.~\ref{fig1}).
Thus, purely hydrodynamic collapse simulations of the latter clouds (e.g.,
\markcite{hun77}Hunter 1977; \markcite{fos93}Foster \& Chevalier 1993) differ
significantly from the results of \markcite{shu77}Shu (1977). The approach to
an $r^{-2}$ profile at $t=0$ is rather violent for them, and involves
appreciable fluid velocities that invalidate the main premise ($u=0$ at $t=0$)
of the similarity solutions discussed here and by \markcite{shu77}Shu.  The
subsequent $t>0$ collapse then proceeds much more rapidly than for a singular
sphere, with an accretion rate that decreases over time. In our case, the
logotropic singular solution ($\rho\propto r^{-1}$:  eq.~[\ref{eq:20}] and
Fig.~\ref{fig1}) {\it does} describe the structure of bounded clouds of any
interesting mass. The adjustment to a fully singular profile at $t=0$ should
then be a relatively easy one, and it is reasonable to expect that equations
(\ref{eq:213}) adequately describe any collapsing logotrope at the point of
core formation.  This argument is consistent with numerical simulations
(\markcite{fos93}Foster \& Chevalier 1993) for the collapse of a bounded
isothermal sphere whose central concentration is $\sim$70 times that of the
Bonnor-Ebert value. Although such a cloud is highly unstable, it is well
approximated at large radii by the singular isothermal sphere, and its
accretion rate quickly approaches the constant value predicted by
\markcite{shu77}Shu (1977).

Turning now to the opposite limit of small radii, $x\rightarrow0$, we have
that $|v|,\ \alpha\rightarrow\infty$ for collapse, and thus
\begin{equation}
\alpha\rightarrow4\left({{m_0}\over{2x^3}}\right)^{1/2}\ \ \ \ \ \ 
v\rightarrow -\left({{2m_0}\over{x}}\right)^{1/2}\ \ \ \ \ \ 
m\rightarrow m_0\ ,
\label{eq:214}
\end{equation}
where the reduced mass $m_0$ of the collapsed core also serves to fix the
accretion rate onto the core ($\dot{M}_0\propto m_0t^3$ by eq.~[\ref{eq:27}];
see also \S3 below). The effects of shocks (e.g., \markcite{tsa95}Tsai \& Hsu
1995) and radiation pressure (for massive stars; \markcite{jij96}Jijina \&
Adams 1996) will invalidate this solution at very small $x$, but should not
affect our discussion of the area external to such disturbed regions, nor the
value of $m_0$ itself.

The $x^{-3/2}$ and $x^{-1/2}$ scalings of $\alpha$ and $v$ here are identical
to the corresponding results for the isothermal sphere (\markcite{shu77}Shu
1977), or any other polytrope with $0<\gamma<1$ (Appendix A). This reflects
the fact that the ratio of gravitational and pressure forces,
\begin{equation}
{{GM/r^2}\over{\rho^{-1}dP/dr}}\ =\ 
{{m\alpha}\over{x}}\left({{d\ {\hbox{ln}}\alpha}\over{d\ {\hbox{ln}}x}}
\right)^{-1}\ 
\rightarrow\ -{{4\sqrt{2}}\over{3}}\ {{m_0^{3/2}}\over{x^{5/2}}}\ ,
\label{eq:214a}
\end{equation}
increases without limit as $x\rightarrow0$, and the gas at very small radii
approaches the core in free-fall. At any fixed time $t>0$, the dimensional
velocity field near the center therefore tends to $u=(2GM_0/r)^{1/2}$, and the
density profile to
\begin{equation}
\rho(r)={{\dot{M}}\over{4\pi r^2u}}={{\dot{M}}\over{4\pi(2GM_0)^{1/2}}}\ 
r^{-3/2}\ ,
\label{eq:214b}
\end{equation}
which is equivalent to equation (\ref{eq:214}) for a spatially invariant
accretion rate $\dot{M}$. When written this way, the effects of the gas EOS
are completely contained in the values of $\dot{M}$ and $M_0$ at a given time;
at least in the absence of rotation and magnetic fields, {\it observations of
the shape of the density profiles or velocity fields in the central regions of
a collapsing cloud are insufficient to constrain the underlying equation of
state}. Any data which favor an $r^{-3/2}$ density profile in such objects
(e.g., \markcite{but90}Butner et al.~1990, \markcite{but91}1991) are
qualitatively consistent with both the logotropic hypothesis and the
isothermal one (quantitatively, however, the different accretion {\it
histories} in the two models may well allow for an empirical discrimination
between them; see \S4 below). By the same token, observations which suggest
a different structure (e.g., \markcite{and93}Andr\'e et al.~1993) are a
challenge to both scenarios.

A full similarity solution for the collapse of overdense clouds, over all
times $t\ge0$ and all radii $r\ge0$, begins at large $x$ as in equation
(\ref{eq:213}), and eventually tends to the form (\ref{eq:214}) at small $x$.
As such, there is a one-to-one relation between $C$ and $m_0$. This is
obtained by numerical integration of equations (\ref{eq:29}) and
(\ref{eq:210}), and is laid out in Table \ref{tab1}. Evidently, the larger the
overdensity $C/\sqrt{2}$, the larger is the reduced core mass $m_0$, and the
faster the accretion rate. The limit $C\rightarrow\sqrt{2}$ refers to the
collapse of a logotrope which is initially arbitrarily close to hydrostatic
equilibrium. In this situation --- which is the parallel of
\markcite{shu77}Shu's (1977) expansion-wave solution for the singular
isothermal sphere --- a collapse front propagates outwards in a cloud at rest,
with the material behind it falling away to the center.

\placetable{tab1}

\subsection{Critical Points and the Expansion Wave}

A look at equations (\ref{eq:29}) and (\ref{eq:210}) shows that when a
collapse solution passes through a point $x_*\ge0$ which has
\begin{equation}
2x_*-v_*-\alpha_*^{-1/2}=0\ ,
\label{eq:215}
\end{equation}
the derivatives $d\alpha/dx$ and $dv/dx$ will be undefined unless the
quantities on the right-hand side of the fluid equations also vanish there.
Any valid flow must therefore satisfy
\begin{equation}
{{\alpha_*^{1/2}}\over{4}}+{1\over{\alpha_*^{1/2}}}-{2\over{x_*\alpha_*}}
+2x_*=0
\label{eq:216}
\end{equation}
at such {\it critical points}. Mathematically, the $\alpha(x)$ and $v(x)$
which everywhere satisfy these two relations would present a third analytical
similarity solution for our problem; but it is one which relies on a rather
special interplay between the fluid variables, and one which we therefore
neglect. Instead, physical flows must either avoid the line of critical points
(or singular points, or sonic points) defined by equations (\ref{eq:215}) and
(\ref{eq:216}), or else pass right through it.

\markcite{whi85}Whitworth \& Summers (1985) elaborate on the topology of
critical points and the implications for a self-similar, isothermal collapse,
and we refer the interested reader to that paper for an excellent discussion
of such issues. Here we note that, for any gas EOS, a linearization of the
fluid equations in the neighborhood of any critical point $x_*$ leads to three
pairs of eigenvalues for the slopes $d\alpha/dx$ and $dv/dx$ there, and hence
to three eigensolutions for $\alpha(x)$ and $v(x)$. One of these is physically
inadmissible but the other two, corresponding to the ``minus'' and ``plus''
solutions of \markcite{shu77}Shu (1977), have the velocity either decreasing
or increasing towards smaller $x$. They are described quantitatively, for
the logotrope and for negative-index polytropes, in Appendix B.

It is the minus solutions, for which $v\rightarrow-\infty$ as $x\rightarrow0$,
that are of most interest in the collapse context. The velocity fields for
some representative examples are shown as the solid lines in
Fig.~\ref{fig2}.  Also shown there are the locus of critical points (dash-dot
line) defined above, and a few plus solutions (dashed lines).  We will not
discuss the plus solutions, other than to say that (1) their time-reversed
counterparts could be of interest in a wind problem, and (2) in the limit
$x\rightarrow\infty$, they have $\alpha\sim x^{-4}$, $v\sim x^{2}$, and
$\alpha v^2=\rho u^2/AP_c=2$.

\placefigure{fig2}

Our Fig.~\ref{fig2} should be compared to that of \markcite{shu77}Shu (1977).
The heavy black line here is the expansion-wave solution mentioned above.  It
passes through the critical point $x_*=0.02439$, and is the limit of two
different sequences of minus solutions. Those which pass through critical
points $x_*<0.02439$, to the left of the expansion wave, go on to a stagnation
point ($v=0$) at some finite $x$, beyond which there is outflow. These
solutions are therefore of no interest to us, although they might be useful in
modelling the effects of accretion shocks (see \markcite{tsa95}Tsai \& Hsu
1995). If they were to be simply truncated at the point where $v=0$, they
would have vanishingly small spatial extent at $t=0$, which is unacceptable;
and, as in the isothermal case, the static solution (eq.~[\ref{eq:212}])
cannot be used to continue such flows to infinite $x$ because there is a
mismatch in the densities at the stagnation point. The densities and
velocities still follow the power laws of equation (\ref{eq:214}) at small
$x$, though, with a core mass $m_0$ that is uniquely defined by the location
of the critical point $x_*$ and is always less than the expansion-wave value.

Fundamentally, the expansion wave is identified as an extreme member of
this family of solutions because its stagnation point is also a critical
point, $x_*= x_{\rm{ew}}$. In this case (see Appendix B),
\begin{equation}
x_{\rm{ew}}={1\over{4\sqrt{2}}}\ \ \ \Rightarrow\ \ \ \alpha_{\rm{ew}}=8
\ \ \ \ \ \ v_{\rm{ew}}=0\ \ \ \ \ \ m_{\rm{ew}}={1\over{32\sqrt{2}}}
\ \ \ \ \ \ m_0=6.67\times10^{-4}\ .
\label{eq:217}
\end{equation}
These results do agree with the static cloud profile (\ref{eq:212}) at
$x=x_{\rm{ew}}$, and the core mass $m_0$ corresponds to that in the
$C\rightarrow\sqrt{2}$ collapse (cf.~Table \ref{tab1} above), so an extension
to arbitrarily large $x$ is obtained by simply joining the two solutions in
the obvious way.  In fact, the logotropic expansion wave comes into
$x_{\rm{ew}}$ from the left with velocity and density derivatives,
$(dv/dx)_{x_{\rm{ew}}}=0$ and $(d\alpha/dx)_{x_{\rm{ew}}}=-1/32\sqrt{2}$,
which also agree with those of the static profile, i.e., the flow is {\it
continuous} at $x_{\rm{ew}}$. This is the case for any polytropic EOS with
$\gamma\le3/5$, but when $\gamma>3/5$ (including the isothermal $\gamma=1$)
there is a discontinuity at the matching point (see Appendix B).

Physically, as discussed above and by \markcite{shu77}Shu (1977), an
expansion wave corresponds to the inside-out collapse of a cloud
which is in hydrostatic equilibrium at the moment $t=0$ of core formation. At
any point in time (before the expansion wave reaches the initial cloud
surface), a total reduced mass of $m_{\rm{ew}}$ is involved in the collapse,
with a small fraction $m_0/m_{\rm{ew}}\simeq0.03$ of that having already
fallen onto the central core. It may also be shown (for any gas EOS) that
since the point $x_{\rm{ew}}$ is fixed in similarity space, the wave
propagates into the cloud always at the {\it local} speed of sound,
$(dP/d\rho)^{1/2}$.

However, the logotrope differs from the isothermal sphere, in that its
expansion wave passes through two critical points: $x_*=0.02439$ and
$x_*=x_{\rm{ew}}$.  (For an isothermal EOS, the expansion-wave $v$ is just
tangent to the locus of critical points at $x_{\rm{ew}}$, and does not cross
it anywhere.) It is the presence of these two distinct critical points, {\it
both} with $v_*\le0$, which also marks the expansion-wave flow as the limit of
a second sequence of minus solutions, namely, the out-of-equilibrium collapse
profiles encountered in \S2.1 (eq.~[\ref{eq:213}]). Several of these are drawn
on Fig.~\ref{fig2}, to the right of the expansion wave, and labelled by their
density normalizations $C$. Evidently, for $C$ close enough to the
hydrostatic-equilibrium value of $\sqrt{2}$, such solutions --- which have
$v<0$ everywhere --- also pass through two critical points. This is again in
contrast to the isothermal situation, in which all out-of-equilibrium
collapses are free of critical points. On some level, however, this is just a
mathematical issue; there is nothing intrinsically unphysical or artificial
about flow through a critical point.\footnotemark
\footnotetext{Fig.~\ref{fig2} shows that the $C\simeq1.4245$ collapse just
brushes the locus of critical points at a single $x_*\simeq0.1034$. Solutions
for $C>1.4245$ then bypass all critical points, while those with
$\sqrt{2}<C\le1.4245$ encounter two critical points (one less than 0.1034, and
one greater). In the language of \markcite{whi85}Whitworth \& Summers (1985),
the points with $x_*<0.1034$ are {\it saddles}; those with $x_*>0.1034$, {\it
nodes}; and $x_*=0.1034$ itself is degenerate.  (The direction of stable
numerical integration is away from a saddle, but towards a node; and only a
pure plus or minus solution can pass through a saddle, while any function can
cross a node.) The degeneracy moves towards larger $x$ for increasing
polytropic index $\gamma$, and occurs at $x_*=1=x_{\rm{ew}}$ in the limit
$\gamma=1$. This is why the isothermal expansion wave is only ever tangent to
the line of critical points, and why the overdense collapse solutions ($C>2$
in that case) do not involve critical points at all.}

\section{Accretion Timescales}

Once numerical solutions to the self-similar fluid equations have been
obtained, and the reduced mass $m_0$ found for any minus solution, we have a
description of the growth of the central core in a collapsing cloud.  The
formation times for stars of various masses can then be computed. Again, we
focus on the results for logotropic expansion waves in this Section.
Generalizations to out-of-equilibrium collapses, and to more general
polytropic equations of state, are fairly straightforward (cf.~Appendix A).

From equations (\ref{eq:28}) for the dimensionless mass $m$ and
(\ref{eq:214}) for our similarity solutions at small radii, the central core
mass is given by
\begin{eqnarray}
M(0,t)\ \equiv\ M_0 & = & [AP_c(4\pi G)]^{3/2}\ {{m_0t^4}\over{G}} \nonumber
\\
 & = & A^{3/2}\left({{3\pi^2}\over{8}}{{\rho_c}\over{\rho_{\rm{ave}}}}
\right)^{3/2}{t^4\over{\overline{t}_{\rm{ff}}^3}}\ {{m_0\sigma_c^3}\over{G}}
\ . \label{eq:31}
\end{eqnarray}
Here $\rho_c$ and $\rho_{\rm{ave}}$ are the central and mean densities of the
initial ($t=-\infty$) cloud; $\overline{t}_{\rm{ff}}$ is its mean free-fall
time, $(3\pi/32G\rho_{\rm{ave}})^{1/2}$; and it will be recalled that $A=0.2$
in the logotropic EOS is most appropriate for real molecular clumps and GMCs.
The ratio $\rho_c/\rho_{\rm{ave}}$ and the free-fall time are referred to
their values in the non-singular cloud because these are then related to the
surface pressure, kinetic temperature, and total mass (by, e.g., Table
\ref{tab0} above), which fundamentally control the collapse. As also discussed
in \S2.1, $\sigma_c$ is just the thermal linewidth, $(kT/\mu m_H)^{1/2}$. The
logotropic EOS already accounts for the presence of turbulence in the initial
cloud, and $\sigma_c$ is therefore {\it not} to be interpreted as the type of
``effective'' sound speed discussed by, e.g., \markcite{shu87}Shu et al.~1987.
The accretion rate is therefore
\begin{equation}
\dot{M}_0\ =\ 4A^{3/2}\left({{3\pi^2}\over{8}}{{\rho_c}\over
{\rho_{\rm{ave}}}}\right)^{3/2}\left({t\over{\overline{t}_{\rm{ff}}}}\right)^3
{{m_0\sigma_c^3}\over{G}}\ =\ 4\langle\dot{M}_0\rangle\ ,
\label{eq:32}
\end{equation}
where $\langle\dot{M}_0\rangle$ is the mean rate $M_0/t$. As we discuss
further in \S4, the time dependence of this quantity is the most important
consequence of the move away from an isothermal EOS, which gives a constant
$\dot{M}_0=\langle\dot{M}_0\rangle=0.975\sigma_c^3/G$ for the expansion wave
(\markcite{shu77}Shu 1977; also Appendix A below). $\langle\dot{M}_0\rangle$
increases with time in our case, and in any negative-index polytrope, because
density profiles shallower than $r^{-2}$ put more mass at larger radii in a
cloud.

Now consider a pressure-truncated logotrope which is in hydrostatic
equilibrium, and large enough that the singular $r^{-1}$ density profile holds
at the edge of the cloud. The analysis of \markcite{mcl96}MP96 gives the total
mass and radius as
\begin{equation}
M_{\rm{tot}}={{(3A)^{3/2}}\over{\pi}}\left({{\rho_c}\over{\rho_{\rm{ave}}}}
\right)^{3/2}\overline{t}_{\rm{ff}}\ {{\sigma_c^3}\over{G}}
\label{eq:33}
\end{equation}
and
\begin{equation}
R={{2(3A)^{1/2}}\over{\pi}}\left({{\rho_c}\over{\rho_{\rm{ave}}}}
\right)^{1/2}\overline{t}_{\rm{ff}}\ \sigma_c\ .
\label{eq:34}
\end{equation}
Apart from corrections for the presence of mean magnetic fields
(\markcite{mcl96}MP96), these expressions are valid for a cloud of any mass
as long as the $r^{-1}$ profile is achieved inside it. In turn, as we have
seen (cf.~Fig.~\ref{fig1}), this occurs for the full range of clump masses
--- from $<1$ to several hundred $M_\odot$ --- seen in GMCs. As the expansion
wave moves outwards in a logotrope, equations (\ref{eq:31}) and (\ref{eq:33})
give for the fractional core mass (independently of the EOS parameter $A$)
\begin{equation}
{{M_0}\over{M_{\rm{tot}}}}={{\pi^4}\over{16\sqrt{2}}}
\left({t\over{\overline{t}_{\rm{ff}}}}\right)^4\ m_0\ ,
\label{eq:35}
\end{equation}
where, again, $m_0=6.67\times10^{-4}$. Also, the definition of $a_t$
(eq.~[\ref{eq:25}]) locates the boundary of the cloud in similarity space:
\begin{equation}
X\equiv{R\over{a_tt}}={{4\sqrt{2}}\over{\pi^2}}
\left({t\over{\overline{t}_{\rm{ff}}}}\right)^{-2}\ .
\label{eq:36}
\end{equation}

Since the head of the expansion wave is at $x_{\rm{ew}}= 1/4\sqrt{2}$ for the
logotrope, equation (\ref{eq:36}) shows that it reaches the surface of a
collapsing cloud (of any mass) at
\begin{equation}
t_{\rm{ew}}={{4\sqrt{2}}\over{\pi}}\ \overline{t}_{\rm{ff}}\ ,
\label{eq:37}
\end{equation}
i.e., after $\simeq$1.80 initial free-fall times. At that point, equation
(\ref{eq:35}) (or the ratio $m_0/m_{\rm{ew}}$ from eq.~[\ref{eq:217}]) gives
$M_0/M_{\rm{tot}}\simeq0.03$. If the similarity solution continues to hold
after this time, then the entire cloud will be accreted onto the core (i.e.,
$M_0/M_{\rm{tot}}=1$ in eq.~[\ref{eq:35}]) by
\begin{equation}
t_{\rm{end}}={2\over{\pi}}\left({{\sqrt{2}}\over{m_0}}\right)^{1/4}
\overline{t}_{\rm{ff}}\simeq4.32\ \overline{t}_{\rm{ff}}\ .
\label{eq:38}
\end{equation}
The inside-out collapse described by the expansion-wave solution is the
gentlest possible, and therefore also the slowest. Solutions of the type
(\ref{eq:213}), with $C>\sqrt{2}$ for a sphere which is initially denser than
allowed for equilibrium, lead to faster accretion because of an increase in
$m_0$ (Table \ref{tab1}) and because of an increase in the total mass
(\ref{eq:33}), by a factor $(C/\sqrt{2})^{3/2}$ (see Appendix A).

Realistically, once the expansion wave reaches the boundary of a cloud which
is under a constant surface pressure, it should be reflected and driven back
into the cloud as a compression wave. Thus, our similarity solution cannot be
exact beyond $t_{\rm{ew}}$, and the resulting estimate of $t_{\rm{end}}$ might
be something of an upper limit. However, it does seem likely that more than
one additional free-fall time is required to complete the collapse after
$t_{\rm{ew}}$, because the accreting gas is in free is in free-fall only deep
inside the expansion wave (at $x<0.1 x_{\rm{ew}}$; cf.~eq.~[\ref{eq:214a}]).
In the absence of detailed numerical simulations, then, the most secure
conclusion is that the collapse of a (near-)equilibrium logotrope lasts
for $t_{\rm{end}}>t_{\rm{ew}}+\overline{t}_{\rm{ff}}\ga3
\overline{t}_{\rm{ff}}$ after core formation. However, if the isothermal
sphere is any guide, then the simulations of \markcite{bos82}Boss \& Black
(1982) and \markcite{fos93}Foster \& Chevalier (1993) suggest that the correct
time to $M_0=M_{\rm{tot}}$ for a logotrope may in fact be fairly close to our
equation (\ref{eq:38}): The core in \markcite{bos82}Boss \& Black's numerical
collapse of a non-rotating, equilibrium singular sphere grows at roughly the
expansion-wave rate even after $t_{\rm{ew}}$; and the results of our Appendix
A can be used to compute a $t_{\rm{end}}$ which is only $\sim$20\% longer than
the total accretion time in \markcite{fos93}Foster \& Chevalier's
``$\xi_{\rm{max}}=40$'' run for a cloud that is overdense by 10\% and has
high but finite concentration at $t=-\infty$.

Figure \ref{fig3} shows the density profile and velocity field, in similarity
space, for a full expansion-wave solution. The thick black lines lie inside
the head of the expansion wave, which is at $x_{\rm{ew}}\simeq0.177$.  At
larger $x$, the density profile is given by the hydrostatic singular solution
$\alpha=\sqrt{2}/x$. This is shown as the thin solid line in the top panel,
where the {\it continuous} matching of densities across $x_{\rm{ew}}$ is
evident. The velocity field also joins continuously to the static $v\equiv0$
at $x_{\rm{ew}}$. Table \ref{tab2} lists $\alpha$, $v$, and the mass $m$ as
functions of $x$ inside the head of the expansion wave. The initial radius $R$
of the cloud is fixed in real space (at least, until the expansion wave
reaches it), and so over time moves steadily inwards in similarity space. The
vertical lines in Fig.~\ref{fig3} are placed at $X=R/a_tt$
(eq.~[\ref{eq:36}]), for the times $t/\overline{t}_{\rm{ff}}$ shown. As
discussed above, $X=x_{\rm{ew}}$ after 1.80 free-fall times. Finally, the
dashed lines in both panels of the Figure are just the free-fall profiles of
equation (\ref{eq:214}). Clearly, they do not describe the flow until the
gravitational attraction of the core exceeds the pressure-gradient force, at
$x<0.01$ or so.

\placetable{tab2}

\placefigure{fig3}

A dimensional version of Fig.~\ref{fig3} is given in \S4 below, but we now
examine the dimensional core masses $M_0$ and accretion rates
$\dot{M}_0=4\langle\dot{M}_0\rangle$ as functions of time in logotropes of
various total masses. These are obtained from equation (\ref{eq:35}):
\begin{eqnarray}
M_0 & = & 5.41\times10^{-6}\ M_\odot\ 
\left({{M_{\rm{tot}}}\over{M_\odot}}\right)\ 
\left({{\overline{t}_{\rm{ff}}}\over{4.8\times10^5\ \hbox{yr}}}\right)^{-4}\ 
\left({t\over{10^5\ \hbox{yr}}}\right)^4
\nonumber \\
\dot{M}_0 & = & 2.16\times10^{-10}\ M_\odot\ {\hbox{yr}}^{-1}\ 
\left({{M_{\rm{tot}}}\over{M_\odot}}\right)\ 
\left({{\overline{t}_{\rm{ff}}}\over{4.8\times10^5\ \hbox{yr}}}\right)^{-4}\ 
\left({t\over{10^5\ {\hbox{yr}}}}\right)^3\ .
\label{eq:311}
\end{eqnarray}
It will be recalled (from Table \ref{tab0} above) that
$\overline{t}_{\rm{ff}}=4.8\times10^5$ yr is appropriate for a solar-mass,
$A=0.2$ logotrope with temperature $T=10$ K and surface pressure
$P_s=1.3\times10^5\ k\ {\hbox{cm}}^{-3}\ {\hbox{K}}$. The free-fall time of a
logotrope with any other $M_{\rm{tot}}$, $P_s$, or $T$ can be computed as
discussed in \S2.1, so that equations (\ref{eq:311}) are quite generally
applicable. The first of these relations is plotted in Fig.~\ref{fig4} for a
number of different initial cloud masses. In a $C>\sqrt{2}$ collapse, our
$M_0$ and $\dot{M}_0$ should each be multiplied by $(C/\sqrt{2})^{3/2}
(m_0/6.67\times10^{-4})$, with $m_0$ taken, for example, from Table \ref{tab1}
above. What was already apparent from equation (\ref{eq:35}) is made more
explicit here: a protostar in a collapsing logotrope gathers fully half of its
mass $M_0$ over the last $1-(0.5)^{1/4}\simeq16\%$ of its accretion phase. By
far the majority of a star's total formation time is spent waiting out a
period of very slow and relatively unproductive infall in its parent cloud.

\placefigure{fig4}

Also shown in Fig.~\ref{fig4} is the line $M_0=0.155\ M_\odot\ (t/10^5\
{\hbox{yr}})(T/10\ {\hbox{K}})^{3/2}$ for a singular isothermal sphere of {\it
any} total mass. There are at least three points of interest in a comparison
between this and the logotropic curves. (1) Any isothermal sphere has the {\it
same} time-invariant accretion rate, while more massive logotropes eventually
accrete much more rapidly than clouds of lower $M_{\rm{tot}}$. This is because
a softer-than-isothermal EOS generally gives rise to larger nonthermal motions
on larger scales, i.e., to higher $\sigma_{\rm{ave}}$ in higher-mass clouds
(cf.~\S2.1). Roughly, then, since $\dot{M}\sim\sigma^3/G$, the mass
originating from larger radii in bigger clouds will fall faster onto the core.
(2) Generally speaking, low-mass stars take longer to accrete in a logotrope
than in an isothermal sphere, while high-mass stars form on relatively shorter
timescales. The extent of the discrepancy between the two models depends on
the initial mass of the logotrope. Indeed, because of the effect noted in
point (1), even the stellar mass for which the total accretion times coincide
depends on $M_{\rm{tot}}$. (3) This being so, there is a much smaller spread
in star-formation times in logotropes than in isothermal spheres. This is due
to the weak mass dependence in the free-fall time of a logotrope (Table
\ref{tab0}), and to the fact that $M_0\propto (t/\overline{t}_{\rm{ff}})^4$.
Specifically, stars of mass $0.3M_\odot\le M_0\le30M_\odot$ all tend to form
within $1-3\times10^6$ yr, in logotropes of essentially any $M_{\rm{tot}}$.
In an isothermal collapse, where $M_0\propto t$, the accretion times would
differ by a factor of 100.

It has to be recalled that these results are based on the assumption that the
self-similar collapse solution continues to hold even after the expansion wave
has been reflected from the boundary of a cloud. As discussed above,
Fig.~\ref{fig4} may then represent upper limits to the true star-formation
times in logotropes. On the other hand, this effect may not be very large; and
in any case, it will be offset to some extent by the (outwards) radiation
pressure from stars larger than a few $M_\odot$ (\markcite{jij96}Jijina \&
Adams 1996). Our analysis should then be a reasonably accurate first-order
approximation to a rather complicated situation.

These points could have some bearing on the formation of bound clusters, which
requires that a large fraction ($\ga$30\%: \markcite{hil80}Hills 1980;
\markcite{lad84}Lada, Margulis, \& Dearborn 1984; \markcite{elm85}Elmegreen \&
Clemens 1985) of the protocluster gas be converted to stars before the
remainder is dispersed by supernovae or \ion{H}{2} regions. In light of
Fig.~\ref{fig4}, and the short accretion times of high-mass stars, this would
seem to demand that high-mass stars {\it start} collapsing {\it after} the
low-mass stars do --- a conclusion which also has ramifications for the form
and origin of the stellar IMF.

\markcite{mye92}Myers \& Fuller (1992; also \markcite{mye93}Myers \& Fuller
1993; \markcite{cas95}Caselli \& Myers 1995) have introduced a ``TNT''
(thermal plus nonthermal) model of molecular clumps, which has an EOS that is
explicitly isothermal at small radii, but becomes essentially polytropic (with
$\gamma<1$) at large $r$. They use this model to estimate approximate
accretion times for stars of various masses. Although our logotropic model is
preferable to the TNT construction (because the latter has density
singularities at the centers of even pre-collapse, $t=-\infty$ clouds), the
$M_0$ vs.~$t$ relation of equation (\ref{eq:311}) and Fig.~\ref{fig4} may
profitably be compared with that of \markcite{mye92}Myers \& Fuller. In
particular, although our $t_{\rm{end}}$ for clouds of various masses tend to
be somewhat longer than the TNT estimates --- which bypass an exact treatment
of the expansion wave, and which are almost isothermal for low-mass clumps
anyway --- they have roughly the same order of magnitude.  More importantly,
the rather narrow range in accretion times, and the fairly rapid formation of
high-mass stars, are common to both sets of results; indeed, these are generic
features of any softer-than-isothermal EOS.

\section{Implications for Low-Mass Star Formation}

Our analysis has concentrated on spherical infall in a logotrope, and ignored
the effects of {\it mean} magnetic fields (to whatever extent disordered
fields contribute to the turbulence in interstellar clouds, they are already
accounted for by the EOS), rotation and circumstellar disks, and radiation
pressure from the forming star. All of these have been studied analytically or
numerically for accretion in isothermal clouds (e.g., Galli \& Shu
\markcite{ga93a}1993a, \markcite{ga93b}b; \markcite{ter84}Tereby et al.~1984;
\markcite{bos82}Boss \& Black 1982; \markcite{jij96}Jijina \&
Adams 1996, and references therein). In addition, some connection has to be
made with the issues of fragmentation and binary formation (see, e.g.,
\markcite{myh92}Myhill \& Kaula 1992).  Clearly, the present study can only be
viewed as one step towards a full theory of non-isothermal collapse.

Nevertheless, some fairly robust conclusions can be drawn. In particular,
although others have considered the effects of softer-than-isothermal, and
even logotropic, equations of state on massive-star formation (e.g.,
\markcite{mye92}Myers \& Fuller 1992; \markcite{jij96}Jijina \& Adams 1996),
the most important new point to be made here is that even low-mass
($\sim1M_\odot$) molecular clumps can be described by a logotropic EOS. Their
collapse is therefore rather gentler than in the isothermal case, and their
accretion rates must be much smaller at early times.

Figure \ref{fig5} shows the time evolution, from $t=5\times10^4$ to
$t=8.6\times 10^5$ yr, of the density profile and velocity field in an
expansion wave moving through an initially hydrostatic $A=0.2$ logotrope with
a total mass of $1M_\odot$, a kinetic temperature $T=10$ K (hence
$\sigma_c=1.88\times10^4$ cm s$^{-1}$), and a constant surface pressure
$P_s=1.3\times10^5\ k$ cm$^{-3}$ K. The velocity scale $a_t$
(eq.~[\ref{eq:25}]) can be evaluated, at any time, given the surface pressure
and the fact (cf.~Table \ref{tab0}) that $P_c/P_s=1.58$ in the non-singular
cloud at $t=-\infty$; the dimensional $r$, $\rho$, and $u$ of Fig.~\ref{fig5}
then follow immediately from the $x$, $\alpha$, and $v$ of Fig.~\ref{fig3}. At
the point $t=0$ of core formation, the cloud is everywhere at rest and follows
the singular density profile $\rho(r)=1.3\times10^{-20}\ {\hbox{g}}\
{\hbox{cm}}^{-3}\ (r/R)^{-1}$. At later times, this profile still holds
outside of the expansion wave; it is sketched as the broken line in the top
panel of Fig.~\ref{fig5}. For reference, we also have that the initial
free-fall time of this cloud is $\overline{t}_{\rm{ff}}=4.8\times10^5$ yr
(Table \ref{tab0}); the initial ratio of central and mean densities is
$\rho_c/\rho_{\rm{ave}}=4.20$; and the initial radius is $R=2.9\times 10^{17}$
cm (eq.~[\ref{eq:34}]; this is within a factor 2 of the radius of a solar-mass
isothermal sphere under the same surface pressure). We have only plotted the
densities and velocities for $t\le8.6\times 10^5\ {\hbox{yr}}= 1.80\
\overline{t}_{\rm{ff}}$, i.e., prior to the time $t_{\rm{ew}}$ when the head
of the expansion wave meets the initial boundary of the cloud.

\placefigure{fig5}

A cloud which is initially rotating at a constant rate $\Omega$ will depart
from the spherical expansion-wave solution at small $r\la R_C$, where $R_C$ is
the {\it centrifugal radius} at which the circular speed $(GM_0/r)^{1/2}$ just
equals the infall velocity $-u$. For an $M_{\rm{tot}}=1M_\odot$ logotrope with
the surface pressure and temperature we have assigned here, it is given by
\begin{equation}
R_C={{\Omega^2M_0}\over{2\pi AP_c}} = 3.0\times10^{10}\ {\hbox{cm}}\
\left({{\Omega}\over{10^{-14}\ {\hbox{s}}^{-1}}}\right)^2
\left({t\over{10^5\ {\hbox{yr}}}}\right)^4\ .
\label{eq:41}
\end{equation}
The first of these equalities comes from \markcite{jij96}Jijina \& Adams
(1996; note the slightly different notation); the second, from our equation
(\ref{eq:311}), along with $A=0.2$ and the fact that $P_c=1.58P_s$ for this
specific example. Thus, a disk of 100 AU in extent will be built up within
$\sim$$1.5\times10^6$ yr in our model, as compared to $\simeq$$7.5\times10^5$
yr for a singular isothermal sphere at $T=10$ K. Scaling of this result for
clouds with different total masses, surface pressures, and temperatures
involves a change in the coefficient of $M_0$, as described after equation
(\ref{eq:311}), and a change in $P_c/P_s$ from, e.g., Table \ref{tab0}. In any
case, the magnitude of $R_C$ at a specified age $t$ gives some indication of
the radial range in which the results of Fig.~\ref{fig5} are valid in an
initially rotating cloud.  Calculation of the analogous length scale for a
magnetized sphere --- the so-called magnetic radius of Galli \& Shu
(\markcite{gal93a}1993a, \markcite{gal93b}b) --- will also be important, but
is beyond the scope of this paper.

There are two main points to Fig.~\ref{fig5}. The first is that the densities
far behind the head of the expansion wave are well in excess of the initial
hydrostatic values (see also Fig.~\ref{fig3} above), and are {\it increasing}
with time. This is in stark contrast to an isothermal collapse, where
$\rho(r)$ falls below its equilibrium value and decreases with time at small
radii (cf.~Fig.~3 of \markcite{shu77}Shu 1977). The difference stems in part
from the increasing $\dot{M}_0$ in our case, but more fundamentally from the
fact that the initial $r^{-1}$ density profile of the logotrope is shallower
than the final $r^{-3/2}$ which must obtain near the central core; there is
relatively more mass in the outer regions of the static cloud than in the
collapsing one, so a backlog of sorts develops at small radii. The opposite is
true of a singular isothermal sphere, which begins with a static $r^{-2}$
density profile. More quantitatively, the scaling $\alpha\propto x^{-3/2}$
implies $\rho(r,t)\propto r^{-3/2}t$ for the logotrope, but $\rho(r,t)\propto
r^{-3/2}t^{-1/2}$ for an isothermal EOS.

This situation calls to mind observations of the far-infrared emission of the
NGC 2071 region in Orion B (\markcite{but90}Butner et al.~1990).
Radiative-transport models are consistent with a central star of around
$5M_\odot$ there, and a $\rho\propto r^{-3/2}$ density profile at radii
$r\ge2000$ AU. However, the value of the density at 2000 AU given by
\markcite{but90}Butner et al.'s radiative transport analysis (and also
suggested by independent molecular data) is at least one full order of
magnitude {\it larger} than that expected for an isothermal expansion-wave
collapse. One explanation for this is that 2000 AU is deep inside the
expansion wave for an isothermal collapse age of $\simeq5\times10^5$ yr (which
follows from an assumed $T=35$ K and $M_0=5M_\odot$), such that significant
depletion would have occurred there. There is no such depletion, but rather an
enhancement, in our logotropic model, and these data could therefore provide
an important test of it. However, a proper analysis of NGC 2071 would require
a self-consistent treatment of radiation pressure and rotation, as well as a
reliable feel for the initial total mass involved, and is too detailed to be
undertaken here.

The second point of Fig.~\ref{fig5} is that, naturally, the infall velocity at
any given radius also grows over time, {\it but} at a rate which is quite
different from that in an isothermal sphere. In particular, since $u\propto
(M_0/r)^{1/2}$ at small radii, we have that $u\propto r^{-1/2}t^2$ in a
logotrope, while $u\propto r^{-1/2}t^{1/2}$ in the isothermal case. At any
fixed $r$, the velocities in the two models will agree at some $t$; the
logotropic $u$ must then be smaller than the isothermal one at earlier times,
and larger later on. With $M_{\rm{tot}}=1M_\odot$ specifically, a comparison
of Fig.~\ref{fig5} and \markcite{shu77}Shu's (1977) Fig.~3 shows that we are
firmly in the former, lower-velocity regime at all radii and throughout the
entire duration $t\le t_{\rm{ew}}$.

For example, let us focus on the scale $r=10^{16}\ {\hbox{cm}}\simeq0.003$ pc.
Our model gives infall velocities there (and at larger radii) which are {\it
less than the thermal linewidth}, $\sigma_c\simeq0.2$ km s$^{-1}$, for all
$t\le8.6\times10^5$ yr. By contrast, a solar-mass, singular isothermal
sphere attains $-u\simeq\sigma_c$ at the same radius by $t=6\times10^4$ yr.
This is one example of how the systematic velocity field at observable radii
in a logotrope of {\it any} mass will be effectively swamped by random
velocities for a significant fraction of the duration of its collapse; even a
molecular clump which is well into the accretion phase can appear to be in
virial equilibrium. Because the accretion rate increases with time, at any
instant the majority of clumps will be in a fairly inconspicuous evolutionary
state.  This could explain the apparent paucity of obviously collapsing
regions of the ISM. Similarly, for those clouds which do show observable
signatures of collapse, the adoption of a logotropic EOS leads to larger
inferred dynamical ages than an isothermal model would suggest.

Currently the best direct, kinematic evidence for collapse is in the isolated
Bok globule B335, where molecular line profiles appear consistent with infall
velocities $-u\sim1.2$ km s$^{-1}$ at radii $r\sim5\times10^{15}$ cm
(\markcite{zho93}Zhou et al.~1993). Given that $-u=(2GM_0/r)^{1/2}$ deep
inside any expansion wave (which radial scaling is also consistent with the
B335 data), this implies the existence of a central protostar (or star plus
disk) with mass $M_0\simeq0.25M_\odot$. As well, 1.3-mm continuum measurements
of dust emission indicate a circumstellar envelope of about $0.8M_\odot$
(\markcite{bot96}Bontemps et al.~1996). The physical parameters which we have
specified in deriving Fig.~\ref{fig5} therefore seem particularly suited to a
discussion of B335.\footnotemark
\footnotetext{In fact, \markcite{zho90}Zhou et al.~(1990) favor a kinetic
temperature of 13 K for B335. A $T=13$ K logotrope with no mean magnetic field
has a critical mass of $155M_\odot$ and a free-fall time of $6.6\times10^5$
yr, by equations (\ref{eq:39}) and (\ref{eq:40}). Thus, B335 has
$M_{\rm{tot}}/M_{\rm{crit}}\simeq0.0064$, and hence a truncation radius
$R/r_0\simeq1$ from Table \ref{tab0}. This in turn implies
$\overline{t}_{\rm{ff}}/\overline{t}_ {\rm{ff,crit}}\simeq0.74$, or
$\overline{t}_{\rm{ff}}=4.9\times10^5$ yr. Our timescale estimates, which
strictly apply to a $T=10$ K solar-mass logotrope with
$\overline{t}_{\rm{ff}}=4.8\times10^5$ yr, are then relevant to B335 as well.}
In particular, {\it model-independent} considerations suggest that $M_0/
M_{\rm{tot}}\sim0.25$ in this object. If it is undergoing a logotropic
collapse, it must then be older than $t_{\rm{ew}}=8.6\times10^5$ yr, because
our similarity solution gives a much smaller $M_0/M_{\rm{tot}}=0.03$ at that
time. (In addition, Fig.~\ref{fig5} shows that $-u$ is rather less than the
observed value at $r=5\times10^{15}$ cm for all $t\le t_{\rm{ew}}$.) That is
to say, {\it the expansion wave should already have been reflected from the
boundary of B335}.

If B335 is much older than $t_{\rm{ew}}$, then it will no longer follow its
original $r^{-1}$ density profile; the entire cloud should be taken up in the
collapse and the free-fall $\rho\propto r^{-3/2}$ should hold essentially
throughout. In fact, this is consistent with the study of
\markcite{zho90}Zhou et al.~(1990), who conclude that the densities in the
inner regions of the globule are best described by an $r^{-1.5}$ power law.
(As a test of the isothermal collapse theory specifically, they go on to argue
that at large radii the densities are consistent with $\rho\propto r^{-2}$.
However, there is no {\it direct} evidence for this, or for any other power
law but $-1.5$; the data are ``insufficient to determine independently the
parameters of a density distribution with two power-law regimes''
[\markcite{zho90}Zhou et al.~1990].)

Just how old is B335? Equation (\ref{eq:311}), given $M_0=0.25M_\odot$,
predicts a collapse age of $t=1.5\times10^6$ yr (if our fiducial surface
pressure is correct here), and hence a time-averaged accretion rate of
$\langle\dot{M}_0\rangle=M_0/t=1.7\times10^{-7}M_\odot\ {\hbox{yr}}^{-1}$.
These numbers are, moreover, quite robust again any uncertainties in the
initial total mass: even for an unrealistically extreme $M_{\rm{tot}}=10
M_\odot$ (in which case $M_0/M_{\rm{tot}}\sim0.03$ and the expansion wave
would have just reached the edge of the cloud), we find $t\simeq1.2\times10^6$
yr.  By contrast, \markcite{zho93}Zhou et al.~(1993) discuss their data in
terms of the isothermal expansion wave model and estimate $t=1.5\times10^5$
yr, $\dot{M}_0=2.8\times10^{-6}M_\odot\ {\hbox{yr}}^{-1}$, and
$M_0=0.4M_\odot$. It seems inevitable that our model requires an age for B335
which is nearly an order of magnitude older, with a time-averaged accretion
rate that much smaller, than previously thought.

B335 is an example of the low-mass ``Class 0'' sources recently identified by
\markcite{and93}Andr\'e et al.~(1993) and \markcite{and94}Andr\'e \& Montmerle
(1994). These typically have (cold) single-blackbody spectral energy
distributions, but are invisible at short wavelengths $\lambda\la10\ \mu$m.
This is attributed to veiling by a circumstellar envelope whose mass exceeds
that of the central protostar --- a conclusion which is supported by direct,
mm-continuum measurements of the envelope masses --- and therefore implies
that these objects are in the earliest stages of collapse yet to be observed.
Indeed, the theory of isothermal accretion applied to the Class 0 sources
generally implies ages of a few 10$^4$ yr (e.g., \markcite{bar94}Barsony 1994;
\markcite{pud96}Pudritz et al.~1996). Although B335 might then be somewhat
older than average even in the context of isothermal models, it seems likely
that the relative age increase we deduce for it should also be applied to
other Class 0 sources. Note, though, that most of this age increase is
required to accomodate an early period of very gentle infall, with small
central source masses and low accretion luminosities. {\it Most} of a
$1M_\odot$ star is accreted over a period of only $\sim2-3\times 10^5$ yr in a
logotrope.  This is much smaller than the total collapse time
(cf.~Fig.~\ref{fig4}) and is still consistent, for example, with empirical
estimates of the formation time of visible T Tauri stars in the Taurus-Auriga
complex (\markcite{mye83}Myers \& Benson 1983; \markcite{mye87}Myers et
al.~1987).

Finally, if our low mean accretion rate in B335 is typical, then it should
have some bearing on the ``luminosity problem'' of young stellar objects in
Taurus-Auriga. Very briefly, the bolometric luminosities of Class I sources
there (which are embedded in less massive circumstellar envelopes than Class
0's, and hence appear to be in later stages of collapse;
\markcite{and94}Andr\'e \& Montmerle 1994) are generally about an order of
magnitude smaller than expected from number-count arguments
(\markcite{ken90}Kenyon et al.~1990; \markcite{ken94}Kenyon et al.~1994) and
fits to individual spectral energy distributions (\markcite{ken93}Kenyon,
Calvet, \& Hartmann 1993; see also \markcite{ada87}Adams, Lada, \& Shu 1987).
A similarly low luminosity has also been noted for the prototypical Class 0
source VLA 1623 (\markcite{pud96}Pudritz et al.~1996). The problem derives, in
large part, from the isothermal assumption of a time-independent accretion
rate $\dot{M}_0$, and Kenyon et al.~note that it could be alleviated if stars
accumulate most of their mass on timescales which are short compared to the
total duration of their embedded phase. We have seen that such a situation
arises naturally in the collapse of a logotrope. A revision of the assumed EOS
in molecular clouds may therefore play an important part in the eventual
resolution of this issue.

\section{Summary}

\markcite{mcl96}McLaughlin \& Pudritz (1996) have shown that a logotropic
equation of state, $P/P_c=1+A\ {\hbox{ln}}(\rho/\rho_c)$ and $A=0.2$, provides
the most self-consistent description of the internal, equilibrium structure of
both low- and high-mass dense clumps within GMCs, and of the global properties
(Larson's laws) of the giant cloud complexes themselves. We have now used this
model to investigate the collapse of interstellar clouds. There are a number
of consequences for issues of not only high-mass, but also low-mass, star
formation.

We have derived similarity solutions for the collapse of a gaseous logotrope.
These solutions are analogous to those of \markcite{shu77}Shu (1977) for the
collapse of singular isothermal spheres; in particular, an expansion-wave
solution --- the inside-out collapse of a cloud which is initially in virial
equilibrium --- is identified as the one of most physical interest. The
predicted density profile is $\rho\propto r^{-1}$ for gas which has not yet
been reached by the expansion wave, and $\rho\propto r^{-3/2}$ far behind the
collapse front, where infall velocities approach their free-fall values
$-u=(2GM_0/r)^{1/2}$.  These latter scalings are model-independent, as they
follow directly from the presence of a central point mass (protostar) $M_0$.
However, the core accretion rate in a logotrope increases with time (as
$\dot{M}_0\propto t^3$, so that $M_0\propto t^4$), while it is constant for an
isothermal sphere.

This implies that $\dot{M}_0$ in a logotrope will be smaller than the
isothermal value at early times, and larger at late times. Low-mass stars
therefore take longer to form, while high-mass stars are built up much more
rapidly. With $\overline{t}_{\rm{ff}}$ the mean free-fall time of a logotrope
(typically of order $5\times10^5$ yr), our similarity solutions are strictly
valid for $t<1.8\overline{t}_{\rm{ff}}$. At that point the expansion wave
reaches the edge of a cloud --- with only 3\% of the total initial mass having
been accreted onto the core --- and hydrodynamical simulations will likely be
required to exactly describe the later stages of evolution in which most of
the final mass of a star is accreted. Nevertheless, a simple extrapolation of
our present results suggests that if a cloud of any mass were to completely
collapse to a single star, it would do so by $t=4.3\ \overline{t}_{\rm{ff}}$.
A weak dependence of $\overline{t}_{\rm{ff}}$ on mass in logotropes, together
with the result $M_0\propto t^4$, guarantees that the accretion times of stars
of widely disparate masses differ only slightly: stars of anywhere from 0.3 to
$30M_\odot$ are all expected to form within $\sim1-3\times10^6$ yr. This
conclusion, which is similar to that reached in other studies of
non-isothermal collapse, has important implications for cluster formation and
the stellar IMF.

Another clear difference between the logotropic and isothermal models lies in
the evolution of the density at small radii in a cloud: $\rho$ ultimately
increases with time at any given radius in our model, but would decrease in a
singular isothermal sphere. Infall velocities $u$, on the other hand, grow
with time in any collapse model, but the {\it rates} at which they do so are
quite dependent on the assumed equation of state. Early on in its collapse,
the logotrope can have much smaller $u$ than an isothermal sphere. It
therefore takes longer for observable systematic velocities to develop at
observable radii within a cloud, and collapse will initially be harder to
detect directly. This implies that (1) early on in its accretion phase, a
logotrope will effectively appear to still be in virial equilibrium, such that
the number of molecular clumps which are in fact undergoing collapse could be
much larger than previously suspected; and (2) current estimates for the ages
of (low-mass) young stellar objects, which are based on isothermal collapse
models, should be subject to a significant upwards revision. In the Bok
globule B335 specifically, we find an age which is certainly greater than
$8.6\times10^5$ yr, and probably $\sim$$1.5\times10^6$ yr, older by perhaps an
order of magnitude than suggested by the isothermal expansion-wave theory.
This result could have wider implications for the ages of Class 0 protostars
(of which B335 is one) in general.

\acknowledgments

This work was supported in part by the Natural Sciences and Engineering
Research Council of Canada.

\appendix

\section{COLLAPSING POLYTROPES}

\subsection{Similarity Solutions}

In many ways, the isothermal sphere and the logotrope are profitably viewed as
opposite extremes of the family of {\it negative-index polytropes}, which have
$P\propto\rho^{\gamma}$ with $0<\gamma\le1$. Although incomplete as models
of real interstellar clouds, these have been considered, for example, by
\markcite{mal88}Maloney (1988), \markcite{mck95}McKee \& Zweibel (1995), and
\markcite{mcl96}MP96. The case $\gamma=1$ gives an isothermal equation of
state, while most features of the logotrope are recovered by formally setting
$\gamma=0$ in what follows. The ``negative-index'' label refers to standard
polytropic notation, which has $\gamma=1+1/N$.  An isothermal EOS then
corresponds to $N\rightarrow -\infty$; the logotrope, to $N\rightarrow-1$.

To derive similarity solutions for collapsing polytropes, we write the EOS as
\begin{equation}
P=K\rho^{\gamma}\ .
\label{eq:a1}
\end{equation}
The similarity variable $x$, the dimensionless density $\alpha$, the velocity
$v$, and the mass $m$ are all defined as in equations (\ref{eq:26})
and (\ref{eq:27}) above; and since $dP/d\rho=K\gamma\rho^{\gamma-1}$, the
appropriate velocity scale is just
\begin{equation}
a_t=\left[K\gamma(4\pi Gt^2)^{1-\gamma}\right]^{1/2}\ .
\label{eq:a2}
\end{equation}
As discussed in \S2.1, we are concerned with the evolution of a cloud only
{\it after} the time $t=0$ when a core has formed at its center. The density
structure is then singular throughout, and $K$ is a constant in space and
time, with the value $\sigma_c^2\rho_c^{\gamma-1}$ it had in the initial
($t=-\infty$) cloud.

If we write
\begin{equation}
\beta\equiv(2-\gamma)x-v\ ,
\label{eq:a3}
\end{equation}
then equations (\ref{eq:21}) -- (\ref{eq:23}) give the following:
\begin{eqnarray}
\left[\beta^2-\alpha^{\gamma-1}\right]{{d\ {\hbox{ln}}\alpha}\over{dx}} & = &
\left[{\alpha\over{4-3\gamma}}-{2\over{x}}\beta\right]\beta +
(1-\gamma)\left[2\beta+v\right] \label{eq:a5} \\
\left[\beta^2-\alpha^{\gamma-1}\right]{{dv}\over{dx}} & = &
\left[{\alpha\over{4-3\gamma}}\beta-{{2\alpha^{\gamma-1}}\over{x}}\right]\beta
+ (1-\gamma)\left[2\alpha^{\gamma-1}+v\beta\right] \label{eq:a6} \\
m & = & {{\alpha x^2}\over{4-3\gamma}}\beta\ . \label{eq:a4}
\end{eqnarray}
Setting $\gamma=1$ in expressions (\ref{eq:a2}) onwards recovers the results
of \markcite{shu77}Shu (1977) for the isothermal sphere; if instead $\gamma=0$
(but $K\gamma$ is replaced by $AP_c$), then the equations of our
\S2.2 are obtained. No constraint has yet been imposed on the value of
$\gamma$, and in particular it could well be $>1$ at this point. In this
regime, treatment of the case $\gamma=4/3$ evidently requires some care. We do
not discuss this point further, but note that it relates to the instability of
positive-index polytropes with $1<\gamma<4/3$ (e.g.,
\markcite{cha67}Chandrasekhar 1967). \markcite{gol80}Goldreich \& Weber (1980)
have studied the {\it homologous} collapse of a $\gamma=4/3$ polytrope, in
connection with the evolution of degenerate iron cores in supernovae. More
generally, \markcite{sut88}Suto \& Silk (1988) solve equations for the
self-similar collapse of polytropes with arbitrary indices $\gamma$. The
physical requirement $K={\hbox{constant}}$ identifies our equations
(\ref{eq:a5}) and (\ref{eq:a6}) with equations (15) of \markcite{sut88}Suto \&
Silk, in their ``$n=2-\gamma$'' case.

Still without restricting the value of $\gamma$, equations (\ref{eq:a5}) --
(\ref{eq:a4}) admit two analytic solutions. One of these is the same
uniform-density case given by equation (\ref{eq:211}); the other is the
static solution,
\begin{equation}
\alpha=C_{\rm{HSE}}\ x^{-2/(2-\gamma)}\ \ \ \ \ \ v\equiv0\ \ \ \ \ \ 
m={{2-\gamma}\over{4-3\gamma}}\ C_{\rm{HSE}}\ x^{(4-3\gamma)/(2-\gamma)}\ ,
\label{eq:a7}
\end{equation}
where the density normalization is
\begin{equation}
C_{\rm{HSE}}=\left[{{2(4-3\gamma)}\over{(2-\gamma)^2}}\right]^{1/(2-\gamma)}
\ .
\label{eq:a10}
\end{equation}
The singular density profile thus obtained is an exact solution of the
equation of hydrostatic equilibrium for a self-gravitating polytrope of any
(negative) index (cf.~\markcite{mcl96}MP96). Again, if $\gamma=0$ here, the
logotropic result (\ref{eq:212}) obtains, while if $\gamma=1$ we have
$\alpha=2/x^2$ and $m=2x$, in agreement with \markcite{shu77}Shu (1977) for
the singular isothermal sphere.

In the context of interstellar clouds, we are interested only in the range
$0<\gamma\le1$ --- and particularly the $\gamma=0$ limit of the logotrope ---
since the velocity dispersion $P/\rho$ then increases with radius. For such
$\gamma$, equations (\ref{eq:a5}) -- (\ref{eq:a4}) in the limit
$x\rightarrow0$ (late times or small radii) and $|v|\gg x$ give
\begin{equation}
\alpha\rightarrow (4-3\gamma)\left({{m_0}\over{2x^3}}\right)^{1/2}
\ \ \ \ \ \ 
v\rightarrow -\left({{2m_0}\over{x}}\right)^{1/2}
\ \ \ \ \ \ 
m\rightarrow m_0\ ,
\label{eq:a8}
\end{equation}
which are minus solutions. In the opposite limit $x\rightarrow\infty$ (early
times or large radii) and $|v|\ll x$, we have instead
\begin{equation}
\alpha\rightarrow Cx^{-2/(2-\gamma)}
\ \ \ \ \ \ 
v\rightarrow -{{2-\gamma}\over{4-3\gamma}}\ C
\left[1-\left({{C_{\rm{HSE}}}\over{C}}\right)^{2-\gamma}\right]\ 
x^{-\gamma/(2-\gamma)}
\ \ \ \ \ \ 
m\rightarrow {{2-\gamma}\over{4-3\gamma}}\ Cx^{(4-3\gamma)/(2-\gamma)}
\ .
\label{eq:a9}
\end{equation}
The density normalization $C$ is essentially a free parameter, but it must
exceed the hydrostatic equilibrium value $C_{\rm{HSE}}$ (eq.~[\ref{eq:a10}])
for inflow ($v<0$). As usual, then, this family of solutions represents the
collapse of non-equilibrium spheres which are everywhere denser (by a factor
$C/C_{\rm{HSE}}$) than a virialized one with the same volume.  There is also a
one-to-one relationship between $C$ and $m_0$ for a complete solution which
extends from $x=\infty$ (the moment of core formation) to $x=0$ (the end of
accretion), and the limit $C\rightarrow C_{\rm{HSE}}^{+}$ corresponds to the
expansion-wave solution. (Note that these asymptotic solutions differ from
those cited by \markcite{lar69}Larson [1969], which hold for plus solutions
where $|v|$ increases at large $x$ but is still $\ll x$. A sufficient, though
not necessary, condition for the existence of such solutions is $\gamma>1/3$.)

\subsection{Accretion Rates}

The reduced core mass $m_0$ translates, by equations (\ref{eq:a2}) and
(\ref{eq:27}), to a dimensional
\begin{equation}
M_0=\gamma^{3/2}\left({{3\pi^2}\over{8}}{{\rho_c}\over{\rho_{\rm{ave}}}}
\right)^{3(1-\gamma)/2}{{t^{4-3\gamma}}\over{\overline{t}_{\rm{ff}}^
{3-3\gamma}}}\ {{m_0\sigma_c^3}\over{G}}\ ,
\label{eq:a11}
\end{equation}
where $\overline{t}_{\rm{ff}}$ is the free-fall time evaluated at the initial
mean density $\rho_{\rm{ave}}$, and where we have used the fact that
$K=\sigma_c^2\rho_c^{1-\gamma}$. By construction, the linewidth $\sigma_c$
at the center of a cloud is taken to be purely thermal --- $\sigma_c^2=kT/\mu
m_H$ --- and the EOS automatically accounts for nonthermal motions elsewhere.
In any case, the mass accretion rate is therefore
\begin{equation}
\dot{M}_0=(4-3\gamma)\gamma^{3/2}\left({{3\pi^2}\over{8}}{{\rho_c}\over
{\rho_{\rm{ave}}}}\right)^{3(1-\gamma)/2}\left({t\over
{\overline{t}_{\rm{ff}}}}\right)^{3(1-\gamma)}{{m_0\sigma_c^3}\over{G}}=
(4-3\gamma)\langle\dot{M}_0\rangle\ ,
\label{eq:a12}
\end{equation}
and only the singular isothermal sphere has an accretion rate $\dot{M}_0=
\langle\dot{M}_0\rangle=m_0\sigma_c^3/G$ that is constant in time. Formally,
the results of \S3 above are recovered by substituting $A^{3/2}$ for
$\gamma^{3/2}$ and setting $\gamma=0$ elsewhere in these expressions.

Now, by manipulating the results given in Appendix C of \markcite{mcl96}MP96,
the total mass and radius of a {\it singular} pressure-truncated polytrope can
be written as
\begin{equation}
M_{\rm{tot}}={8\over{\pi}}\left({{3\gamma}\over{4-3\gamma}}\right)^{3/2}
\left({{4-3\gamma}\over{6-3\gamma}}\right)^{3\gamma/2}
\left({C\over{C_{\rm{HSE}}}}\right)^{3/2}
\left({{\rho_c}\over{\rho_{\rm{ave}}}}\right)^{3(1-\gamma)/2}
\overline{t}_{\rm{ff}}{{\sigma_c^3}\over{G}}
\label{eq:a13}
\end{equation}
and
\begin{equation}
R={4\over{\pi}}\left({{3\gamma}\over{4-3\gamma}}\right)^{1/2}
\left({{4-3\gamma}\over{6-3\gamma}}\right)^{\gamma/2}
\left({C\over{C_{\rm{HSE}}}}\right)^{1/2}
\left({{\rho_c}\over{\rho_{\rm{ave}}}}\right)^{(1-\gamma)/2}
\overline{t}_{\rm{ff}}\sigma_c\ ,
\label{eq:a14}
\end{equation}
which should be compared to equations (\ref{eq:33}), (\ref{eq:34}) in the
logotropic limit $\gamma=0$. The factor $C/C_{\rm{HSE}}$ here is required for
consideration of the collapse (\ref{eq:a9}) of a sphere which is everywhere
denser than a virial-equilibrium configuration with the same radius $R$ and
temperature $\sigma_c$. Where the expansion wave is concerned, of course, the
initial cloud is taken to be exactly virialized, and $C=C_{\rm{HSE}}$.

Given these results, the central point mass builds up as
\begin{equation}
{{M_0}\over{M_{\rm{tot}}}}={\pi\over{8}}\left({{\pi^2}\over{8}}
{{4-3\gamma}\over{2-\gamma}}\right)^{3(1-\gamma)/2}\left({1\over{2-\gamma}}
{C\over{C_{\rm{HSE}}}}\right)^{-3/2}\left({t\over{\overline{t}_{\rm{ff}}}}
\right)^{4-3\gamma}m_0\ ,
\label{eq:a15}
\end{equation}
and the boundary $X=R/a_tt$ is given by
\begin{equation}
X={4\over{\pi}}\left({8\over{\pi^2}}{{2-\gamma}\over{4-3\gamma}}\right)
^{(1-\gamma)/2}\left({1\over{2-\gamma}}{C\over{C_{\rm{HSE}}}}\right)^{1/2}
\left({t\over{\overline{t}_{\rm{ff}}}}\right)^{\gamma-2}\ .
\label{eq:a16}
\end{equation}
If our similarity solution is taken to hold at all times, then equation
(\ref{eq:a15}) gives the time to accrete an entire cloud onto the core
($M_0/M_{\rm{tot}}=1$). For the expansion wave, with $C=C_{\rm{HSE}}$,
this ranges between $t_{\rm{end}}\simeq2.61\ \overline{t}_{\rm{ff}}$ for
$\gamma=1$, and $t_{\rm{end}}\simeq4.32\ \overline{t}_{\rm{ff}}$ for
$\gamma=0$. (The relation between $\gamma$ and $m_0$ is given by Table
\ref{tab3} in Appendix B.) Strictly speaking, however, boundary effects are
important at late times in finite spheres; in particular, the expansion wave
will reach the edge of a cloud when $X=x_{\rm{ew}}$ (eqs.~[\ref{eq:b3}],
[\ref{eq:b4}]), i.e., at
\begin{equation}
t_{\rm{ew}}={{4\sqrt{2-\gamma}}\over{\pi}}\ \overline{t}_{\rm{ff}}<t_{\rm{end}}\ .
\label{eq:a17}
\end{equation}

\section{CRITICAL POINTS}

Collapse flows in polytropes with $\gamma\le1$ (where again $\gamma$ is set to
0 for the logotrope) may involve passage through a critical point, $x_*\ge0$,
where the velocity $v_*\le0$ and density $\alpha_*$ satisfy
\begin{equation}
(2-\gamma)x_*-v_*=\alpha_*^{(\gamma-1)/2}\ .
\label{eq:b1}
\end{equation}
Equations (\ref{eq:a5}) and (\ref{eq:a6}) then demand that
\begin{equation}
{{\alpha_*^{(\gamma+1)/2}}\over{4-3\gamma}}+(1-\gamma)\alpha_*^{(\gamma-1)/2}
-{2\over{x_*}}\alpha_*^{\gamma-1}+(1-\gamma)(2-\gamma)x_*=0
\label{eq:b2}
\end{equation}
in order for $(d\alpha/dx)_{x_*}$ and $(dv/dx)_{x_*}$ to be finite.

The critical point $x_{\rm{ew}}$ for which $v_*=0$ has, by (\ref{eq:b1}),
\begin{equation}
x_{\rm{ew}}={{\alpha_{\rm{ew}}^{(\gamma-1)/2}}\over{2-\gamma}}\ ,
\label{eq:b3}
\end{equation}
so that (\ref{eq:b2}) gives
\begin{equation}
\alpha_{\rm{ew}}=2(4-3\gamma)=\left[{{2(4-3\gamma)}\over{(2-\gamma)^2}}\right]
^{1/(2-\gamma)}x_{\rm{ew}}^{-2/(2-\gamma)}\ ,
\label{eq:b4}
\end{equation}
which agrees with the static solution (eqs.~[\ref{eq:a7}], [\ref{eq:a10}]) of
the fluid equations at $x_{\rm{ew}}$. The collapse solution which begins with
$v\rightarrow-\infty$ at $x=0$, and goes on to pass through $x_{\rm{ew}}$ with
$v=0$, may therefore be continued to larger $x$ by attaching the static
solution. This composite flow is just the expansion wave.

We now confine our attention to critical points $x_*\le x_{\rm{ew}}$, so that
$v_*\le0$ and $\alpha_*\ge\alpha_{\rm{ew}}$. In general, given $x_*$, it is a
simple matter to find $\alpha_*$ and $v_*$ from equations (\ref{eq:b2}) and
(\ref{eq:b1}). We then set
$$\alpha(x)=\alpha_*+a_1(x-x_*)+{\cal O}[(x-x_*)^2]\ \ \ \ \ \ {\hbox{and}}
\ \ \ \ \ \ v(x)=v_*+b_1(x-x_*)+{\cal O}[(x-x_*)^2]$$
in some neighborhood of $x_*$. Substitution of this first-order expansion in
equations (\ref{eq:a5}) and (\ref{eq:a6}) leads to a pair of coupled quadratic
equations for the slopes $a_1$ and $b_1$. Under the constraint (\ref{eq:b2}),
these yield a single cubic equation for $a_1$, and thus three eigensolutions
for the flow through a critical point. The first of these satisfies the simple
relation
\begin{equation}
b_1=(2-\gamma)+{{1-\gamma}\over{2}}a_1\alpha_*^{(\gamma-3)/2}\ ,
\label{eq:b5}
\end{equation}
which serves only to move the flow along the locus (\ref{eq:b1}) of critical
points. Such a propagation of singularities is so contrived as to be
unphysical, and we therefore discard this solution. This leaves us with
\begin{eqnarray}
a_1 & = & - \left[{{k_1\pm\sqrt{k_1^2-4(1+\gamma)k_2}}\over{2(1+\gamma)}}
\right]\ \alpha_*^{(3-\gamma)/2} \nonumber \\
b_1 & = & -2(1-\gamma) + {2\over{x_*}}\alpha_*^{(\gamma-1)/2} +
a_1\alpha_*^{(\gamma-3)/2}\ , \label{eq:b6}
\end{eqnarray}
where
\begin{eqnarray}
k_1 & \equiv & -(9-7\gamma)+{8\over{x_*}}\alpha_*^{(\gamma-1)/2} \nonumber \\
k_2 & \equiv & \alpha_*+2(1-\gamma)(5-3\gamma)-{{4(4-3\gamma)}\over{x_*}}
\alpha_*^{(\gamma-1)/2}+{6\over{x_*^2}}\alpha_*^{\gamma-1}\ . \label{eq:b7}
\end{eqnarray}

For $x_*\le x_{\rm{ew}}$ and $0\le\gamma\le1$, $a_1$ is always real and
negative. If the `$-$' sign is taken to solve for it in equation
(\ref{eq:b6}), we find $b_1\ge0$, and such solutions have $v$ more negative at
smaller $x$. These therefore correspond to the minus solutions of
\markcite{shu77}Shu (1977), or the type 2 solutions of \markcite{hun77}Hunter
(1977) and \markcite{whi85}Whitworth \& Summers (1985). In the isothermal
limit of $\gamma=1$, the result is $a_1=-2/x_*^2$ and $b_1=1/x_*$. If instead
the `$+$' sign is taken in equation (\ref{eq:b6}), then $b_1\le0$ and we
recover the plus (or type 1) solutions; in the isothermal case, $a_1=2/x_*-
6/x_*^2$ and $b_1=1-1/x_*$. The plus solutions for this or any other $\gamma$
have $|v|$ growing for $x>x_*$, and vanishing at some nonzero $x<x_*$. In
addition, they are underdense at large $x$, relative to the singular
hydrostatic profile, and may therefore be of interest as (time-reversed) wind
solutions.

The slopes $a_1$ and $b_1$ at $x_{\rm{ew}}$ can be found analytically, given
any $\gamma$, for both the minus and the plus solutions which flow through
that point; again, the minus solution at $x_{\rm{ew}}$ is the expansion wave.
Using equations (\ref{eq:b3}) and (\ref{eq:b4}) in (\ref{eq:b6}) and
(\ref{eq:b7}), we have the following:
\begin{itemize}
\item $3/5<\gamma\le1$, {\it minus solution}:
\begin{equation}
a_1={{3\gamma-5}\over{1+\gamma}}\alpha_{\rm{ew}}^{(3-\gamma)/2}<0\ \ \ \ \ \ 
{\hbox{and}}\ \ \ \ \ \ b_1={{5\gamma-3}\over{1+\gamma}}>0\ .
\label{eq:b8}
\end{equation}
Although $\alpha_{\rm{ew}}$ and $v_{\rm{ew}}$ match the density and velocity
of the corresponding static solution at $x_{\rm{ew}}$, the derivatives do {\it
not} agree there (in particular, $dv/dx\equiv0$ for the static outer cloud).
The density profile and velocity field of an expansion wave in this range
of $\gamma$ --- which includes the isothermal sphere --- are therefore
discontinuous at $x_{\rm{ew}}$.
\item $0\le\gamma\le3/5$, {\it minus solution}:
\begin{equation}
a_1=-2\alpha_{\rm{ew}}^{(3-\gamma)/2}<0\ \ \ \ \ \ {\hbox{and}}\ \ \ \ \ \ 
b_1=0\ .
\label{eq:b9}
\end{equation}
Now these (left) derivatives do match the (right) derivatives of the static
solution at the head $x_{\rm{ew}}$ of the expansion wave (see
eq.~[\ref{eq:a7}]), and the full flow is everywhere continuous. The logotrope
corresponds to $\gamma=0$, and therefore shows this behavior.
\end{itemize}
The plus solutions at $x_{\rm{ew}}$ for $\gamma>3/5$ have the same $a_1$ and
$b_1$ as the minus solutions for $\gamma\le3/5$; and the plus solutions
at $x_{\rm{ew}}$ for $\gamma\le3/5$ have the same derivatives as the minus
solutions for $\gamma>3/5$.

Equations (\ref{eq:b8}) and (\ref{eq:b9}) also show that, for $\gamma<1$, the
velocity slope $b_1$ of the expansion wave at $x_{\rm{ew}}$ is always
shallower than the gradient $dv_*/dx_*$ of the locus (\ref{eq:b1}) of critical
points there, and the infall velocity of the material just inside the head of
the wave is less than the critical value. However, $|v|$ increases without
bound towards $x=0$ for a minus solution, so any gas EOS which is softer than
isothermal (including logotropic; \S2.3) leads to an expansion-wave flow that
eventually crosses a second critical point $x_*<x_{\rm{ew}}$. Since the
expansion wave also stands as the limit of out-of-equilibrium collapse
solutions ($C>C_{\rm{HSE}}$; Appendix A), some of these must also pass through
two critical points. By contrast, in the isothermal case the expansion wave is
just tangent to the locus of critical points at $x_{\rm{ew}}=1$, and all $C>2$
infall solutions avoid it altogether.

Finally, it is clear that the head $x_{\rm{ew}}$ of the expansion wave; its
second, inner critical point $x_*$; and the core mass/central accretion rate
$m_0$ are all uniquely defined by the specification of the polytropic index
$\gamma$.  These connections are shown in Table \ref{tab3}. Also given there
is the total mass $m_{\rm{ew}}$ interior to $x_{\rm{ew}}$ (from
eq.~[\ref{eq:a7}]). The ratio $m_0/m_{\rm{ew}}$ is the fraction of infalling
material which is already taken up in the central core itself.

\placetable{tab3}

\clearpage

\begin{deluxetable}{lrrlrcc}
\tablecaption{Properties of Pressure-Truncated $A=0.2$ Logotropes.
\label{tab0}}
\tablewidth{0pt}
\tablehead{\colhead{$R/r_0$} &
\colhead{$\rho_c/\rho_s$} & \colhead{$P_c/P_s$} &
\colhead{$\rho_{\rm{ave}}/\rho_c$} &
\colhead{$\sigma_{\rm{ave}}^2/\sigma_c^2$} &
\colhead{$M_{\rm{tot}}/M_\odot$\ \tablenotemark{a}} &
\colhead{$\overline{t}_{\rm{ff}}/10^5$ yr\ \tablenotemark{a}}}
\startdata
~0.93 &   4.28~~ &  1.41~~ & ~0.335  &  2.3~~~ & ~0.5 & 4.2 \nl
~1.34 &   6.22~~ &  1.58~~ & ~0.238  &  2.9~~~ & ~1  & 4.8 \nl
~2.45 &  11.54~~ &  1.96~~ & ~0.130  &  4.4~~~ & ~3  & 5.9 \nl
~4.85 &  23.01~~ &  2.68~~ & ~0.0655  &  6.7~~~ & 10  & 7.1 \nl
~9.40 &  44.63~~ &  4.16~~ & ~0.0337 &  9.1~~~ & 30  & 7.9 \nl
13.18 &  62.56~~ &  5.79~~ & ~0.0240 & 10.0~~~ & 50  & 7.9 \nl
17.02 &  80.77~~ &  8.22~~ & ~0.0186 & 10.1~~~ & 70  & 7.6 \nl
19.26 &  91.39~~ & 10.32~~ & ~0.0164 & 10.0~~~ & 80  & 7.2 \nl
21.26 & 100.88~~ & 12.95~~ & ~0.0149 &  9.3~~~ & 87  & 6.7 \nl
24.37\tablenotemark{b} & 115.58~~ & 20.00~~ & ~0.0130 & 9.0~~~ & 92 & 5.8 \nl
\enddata
\tablenotetext{a}{Calculated assuming no mean magnetic field; surface
pressure $P_s=1.3\times10^5\ k$ cm$^{-3}$ K; and kinetic temperature
$T=10$ K.}
\tablenotetext{b}{Truncation radius of a critically stable cloud,
$M_{\rm{tot}}=M_{\rm{crit}}$.}
\end{deluxetable}

\clearpage

\begin{deluxetable}{ll}
\tablecaption{Core Mass vs.~Density Normalization. \label{tab1}}
\tablewidth{0pt}
\tablehead{\colhead{$C$} & \colhead{$m_0$}}
\startdata
$\sqrt{2}+\epsilon$ & 0.000667 \nl
1.42 & 0.00173 \nl
1.4245 & 0.00248 \nl
1.43 & 0.00339 \nl
1.44 & 0.00507 \nl
1.46 & 0.00869 \nl
1.48 & 0.0127 \nl
1.50 & 0.0170 \nl
1.53 & 0.0243 \nl
1.56 & 0.0323 \nl
1.60 & 0.0444 \nl
1.70 & 0.0808 \nl
1.90 & 0.182 \nl
2.20 & 0.409 \nl
2.60 & 0.871 \nl
\enddata
\end{deluxetable}

\clearpage

\begin{deluxetable}{llllcllll}
\tablecaption{Logotrope Expansion Wave. \label{tab2}}
\tablewidth{0pt}
\tablehead{\colhead{$x$} & \colhead{$\alpha$} & \colhead{$-v$} & \colhead{$m$}
& & \colhead{$x$} & \colhead{$\alpha$} & \colhead{$-v$} & \colhead{$m$}}
\startdata
0     & ~~$\infty$ & ~~$\infty$ & 0.000667 & ~~~~~ & 0.085 & 16.0 & 0.0102 & 0.00522 \nl
0.001 & 2395   & 1.115  & 0.000668 & ~~~~~ & 0.090 & 15.2  & 0.00846  & 0.00582 \nl
0.002 & ~879   & 0.760  & 0.000671 & ~~~~~ & 0.095 & 14.5  & 0.00697  & 0.00646 \nl
0.003 & ~497   & 0.599  & 0.000675 & ~~~~~ & 0.100 & 13.9  & 0.00570  & 0.00713 \nl
0.004 & ~335   & 0.500  & 0.000680 & ~~~~~ & 0.105 & 13.3  & 0.00463  & 0.00784 \nl
0.005 & ~249   & 0.431  & 0.000686 & ~~~~~ & 0.110 & 12.7  & 0.00371  & 0.00859 \nl
0.010 & ~106   & 0.254  & 0.000728 & ~~~~~ & 0.115 & 12.2  & 0.00294  & 0.00938 \nl
0.015 & ~~68.9 & 0.175  & 0.000793 & ~~~~~ & 0.120 & 11.7  & 0.00230  & 0.0102 \nl
0.020 & ~~52.4 & 0.129  & 0.000884 & ~~~~~ & 0.125 & 11.2  & 0.00176  & 0.0111 \nl
0.025 & ~~43.2 & 0.0988 & 0.00100  & ~~~~~ & 0.130 & 10.8  & 0.00132  & 0.0120 \nl
0.030 & ~~37.2 & 0.0781 & 0.00115  & ~~~~~ & 0.135 & 10.4  & 0.000965 & 0.0129 \nl
0.035 & ~~32.9 & 0.0630 & 0.00134  & ~~~~~ & 0.140 & 10.1  & 0.000680 & 0.0139 \nl
0.040 & ~~29.6 & 0.0515 & 0.00156  & ~~~~~ & 0.145 & ~9.74 & 0.000457 & 0.0149 \nl
0.045 & ~~27.0 & 0.0425 & 0.00181  & ~~~~~ & 0.150 & ~9.42 & 0.000289 & 0.0159 \nl
0.050 & ~~24.9 & 0.0354 & 0.00211  & ~~~~~ & 0.155 & ~9.12 & 0.000168 & 0.0170 \nl
0.055 & ~~23.1 & 0.0296 & 0.00244  & ~~~~~ & 0.160 & ~8.84 & 8.50$\times$10$^{-5}$ & 0.0181 \nl
0.060 & ~~21.5 & 0.0248 & 0.00280  & ~~~~~ & 0.165 & ~8.57 & 3.41$\times$10$^{-5}$ & 0.0193 \nl
0.065 & ~~20.2 & 0.0208 & 0.00321  & ~~~~~ & 0.170 & ~8.32 & 8.35$\times$10$^{-6}$ & 0.0204 \nl
0.070 & ~~18.9 & 0.0175 & 0.00366  & ~~~~~ & 0.175 & ~8.08 & 2.86$\times$10$^{-7}$ & 0.0217 \nl
0.075 & ~~17.9 & 0.0146 & 0.00414  & ~~~~~ & $1/4\sqrt{2}$ & ~8 & 0 & $1/32\sqrt{2}$ \nl
0.080 & ~~16.9 & 0.0122 & 0.00466  & & & & & \nl
\enddata
\end{deluxetable}

\clearpage

\begin{deluxetable}{lllll}
\tablecaption{Polytrope Expansion Waves. \label{tab3}}
\tablewidth{0pt}
\tablehead{\colhead{$\gamma$} & \colhead{$x_*$} & \colhead{$x_{\rm{ew}}$} &
\colhead{$m_0$} & \colhead{$m_{\rm{ew}}$}}
\startdata
0   & 0.0244 & 0.17678 & 0.000667 & 0.022097 \nl
0.1 & 0.0340 & 0.21384 & 0.00156  & 0.039756 \nl
0.2 & 0.0472 & 0.25806 & 0.00356  & 0.070502 \nl
0.3 & 0.0658 & 0.31061 & 0.00797  & 0.12287  \nl
0.4 & 0.0917 & 0.37276 & 0.0175   & 0.20964  \nl
0.5 & 0.128  & 0.44583 & 0.0373   & 0.34835  \nl
0.6 & 0.181  & 0.53111 & 0.0774   & 0.55970  \nl
0.7 & 0.259  & 0.62964 & 0.156    & 0.86089  \nl
0.8 & 0.380  & 0.74183 & 0.301    & 1.24939  \nl
0.9 & 0.585  & 0.86668 & 0.558    & 1.67391  \nl
1   & 1      & 1       & 0.975    & 2        \nl
\enddata
\end{deluxetable}

\clearpage

\figcaption[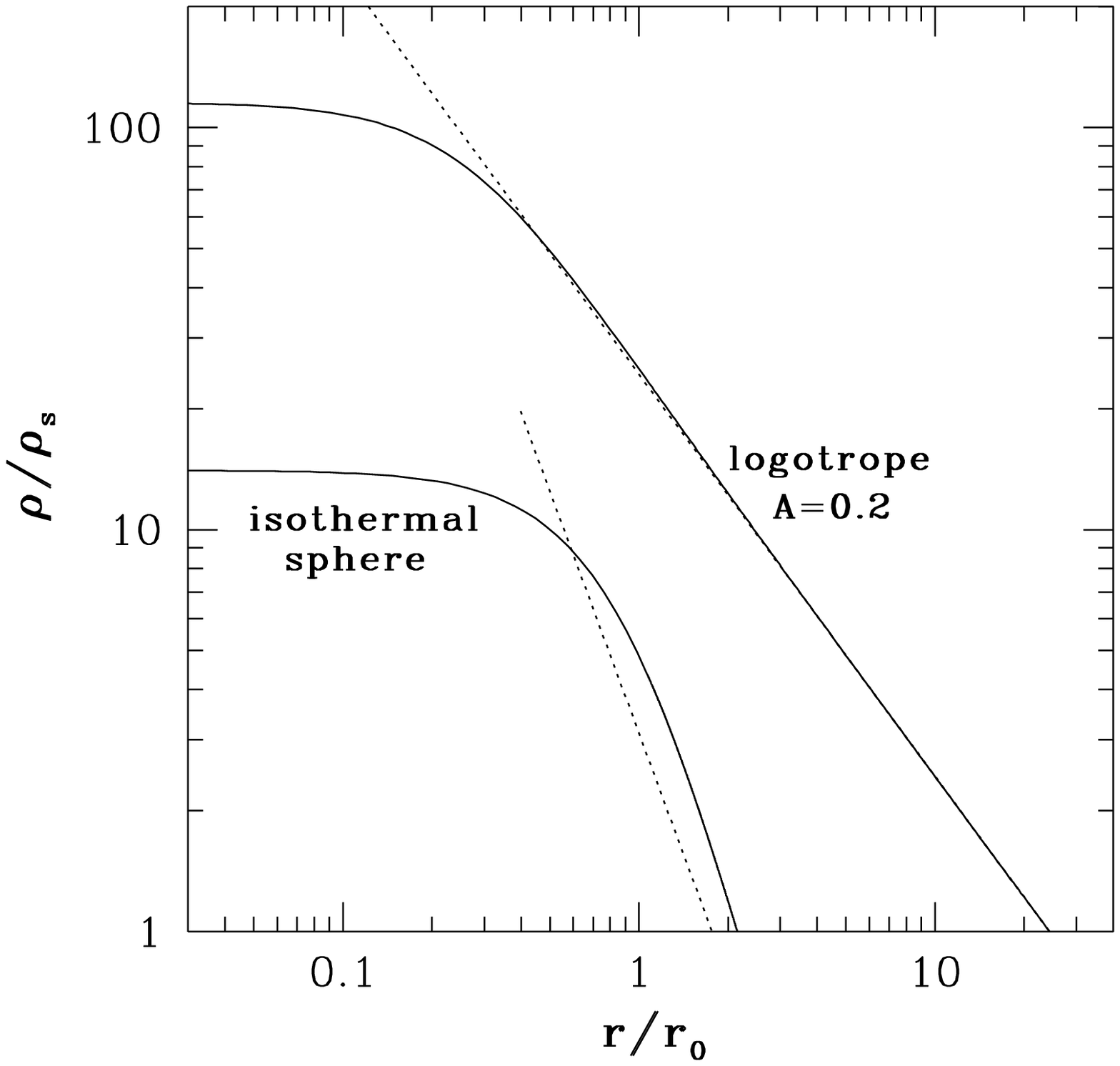]{Density profiles of a pressure-truncated logotrope and
the Bonnor-Ebert isothermal sphere. Both configurations are in hydrostatic
equilibrium, and correspond to critical-mass clouds, i.e., an increased
central concentration in either case would destabilize the equilibrium. The
scale radius is $r_0=3\sigma_c/ (4\pi G\rho_c)^{1/2}$, and $\rho_s$ is the
density at the cloud surface. The broken lines trace the singular profiles for
each equation of state; note how such a solution mirrors the true behavior of
the logotrope at most $r$, but bears little resemblance to any stable
isothermal sphere.
\label{fig1}}

\figcaption[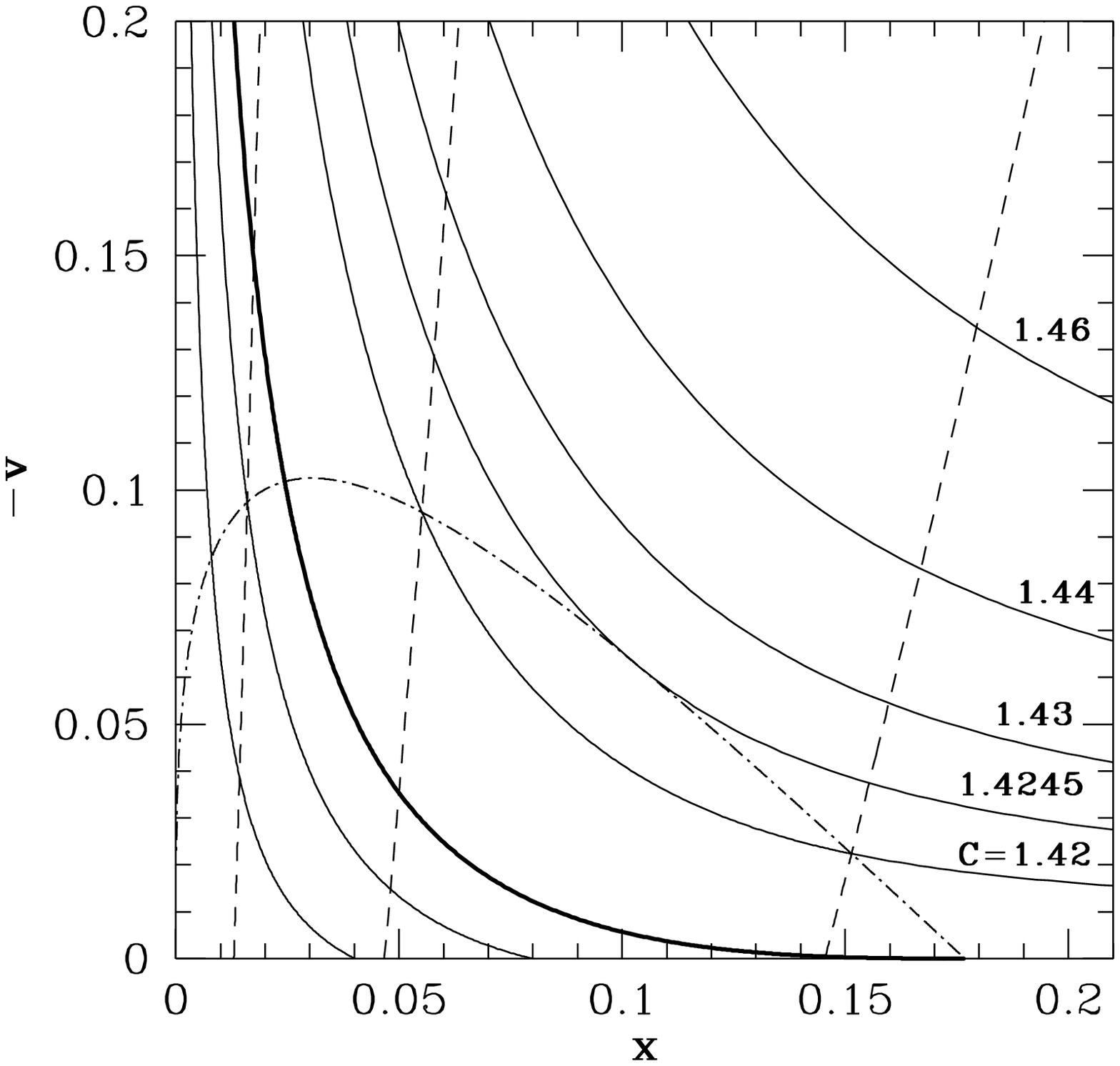]{Velocity fields for various logotrope collapse
solutions. Solid curves are minus solutions, and broken lines are plus
solutions; the dash-dot line traces the locus of critical points. The heavy
solid line represents the expansion wave, which continues to $x=\infty$ with
$v\equiv0$ and densities given by the hydrostatic singular solution. The minus
solutions to the left of the expansion wave reach $v=0$ at some finite $x$,
and therefore have vanishingly small spatial extent at $t=0$. The minus
solutions to the right of the expansion wave follow the collapse of spheres
which are initially out of hydrostatic equilibrium. The expansion wave is the
limit of both sequences.
\label{fig2}}

\figcaption[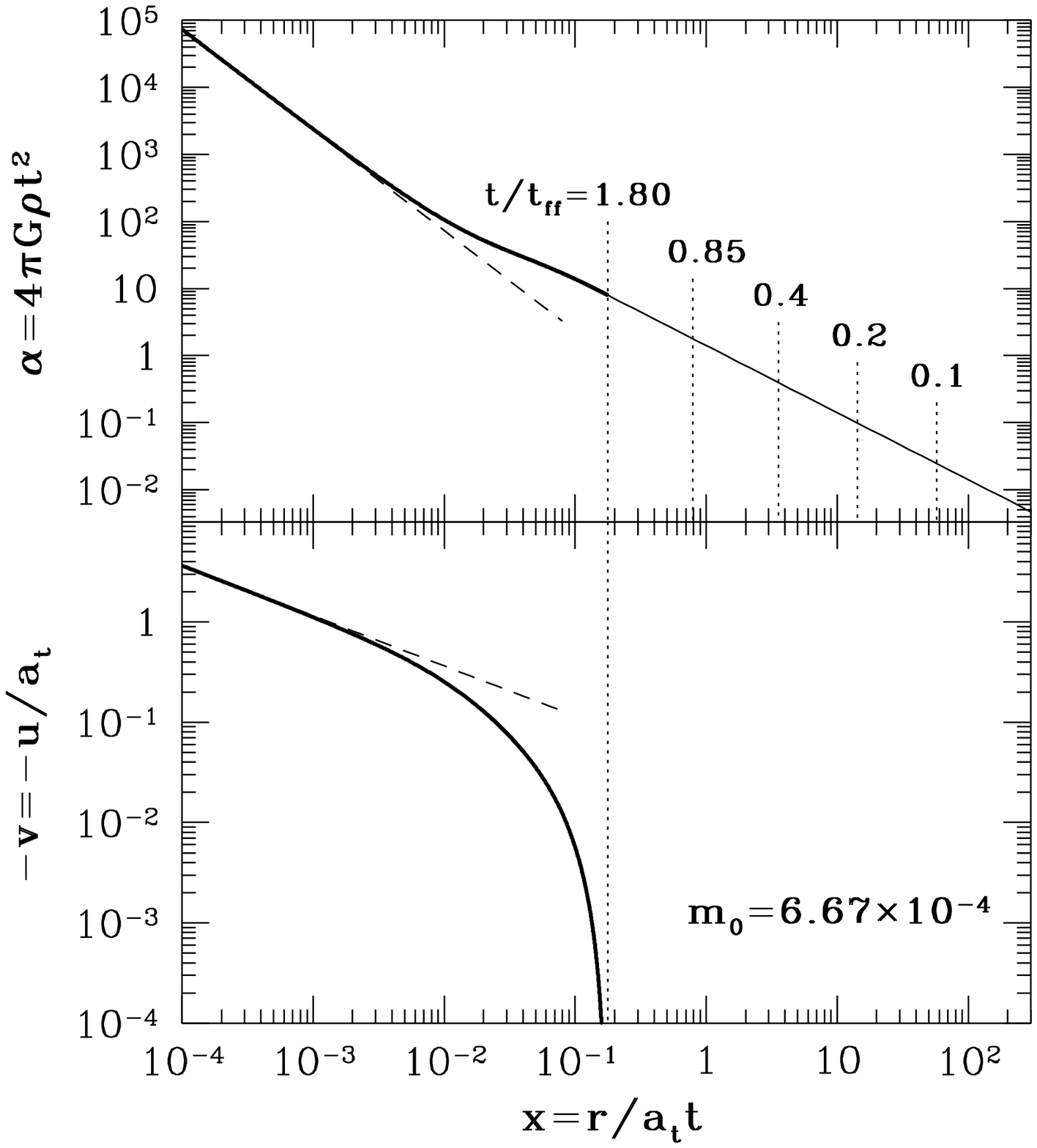]{Density profile and velocity field, in similarity
variables, of the expansion-wave solution for a collapsing logotrope.  The
heavy black lines trace out the region of dynamical collapse inside the
expansion wave, like the same line type in Fig.~\ref{fig2}, while the thin
solid line in the top panel is the hydrostatic singular solution
$\alpha=2^{1/2}/x$. The dashed lines are the expected free-fall solutions
$\alpha\sim x^{-3/2}$ and $v\sim x^{-1/2}$. Vertical (dotted) lines show the
time-dependent location $X=R/a_tt$ of the cloud's outer radius.
\label{fig3}}

\figcaption[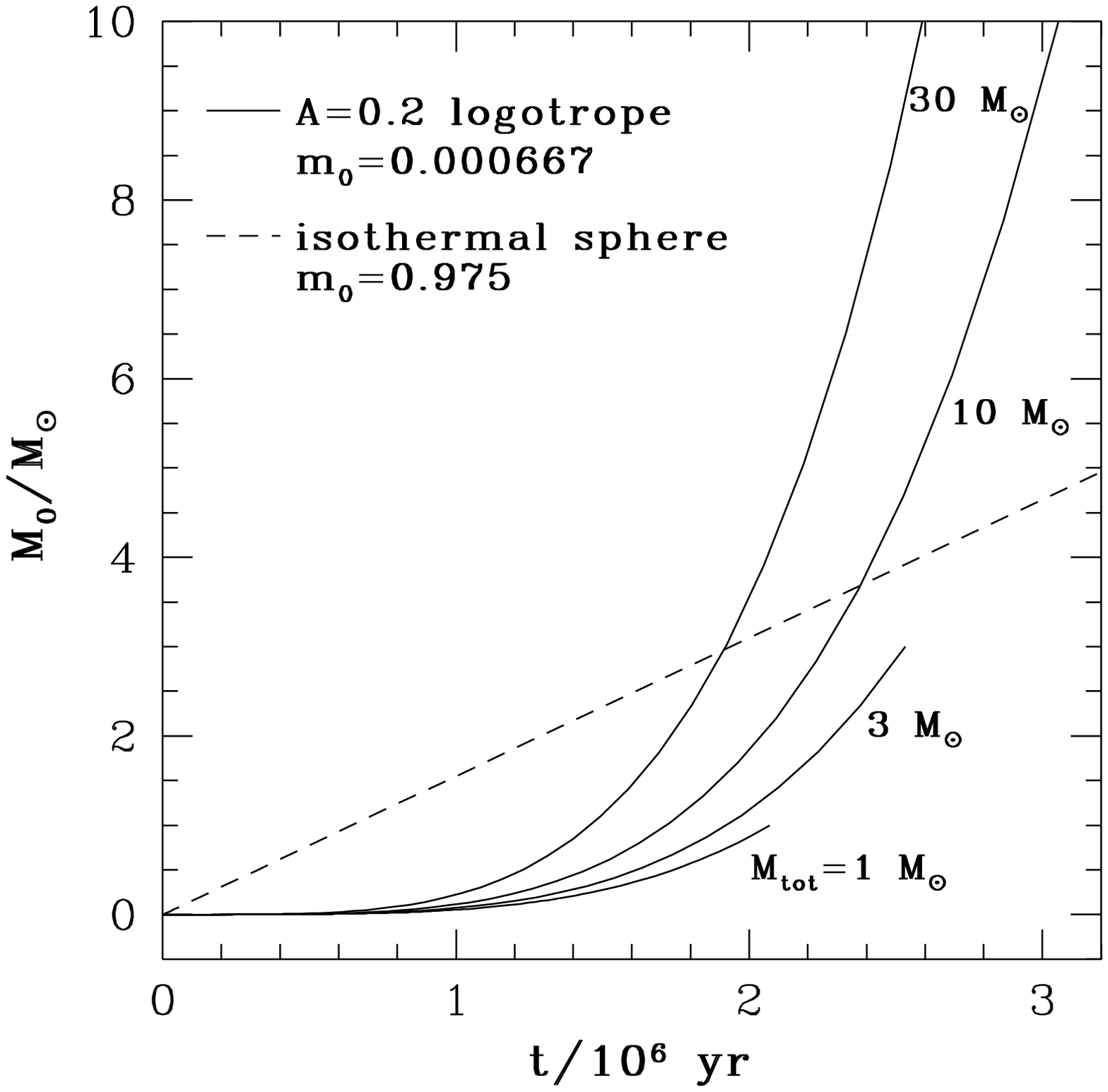]{Expansion-wave accretion timescales for stars in an
$A=0.2$ logotrope and a singular, $T=10$ K isothermal sphere. Low-mass stars
take longer to form in the logotrope, while more massive ones accrete more
rapidly.
\label{fig4}}

\figcaption[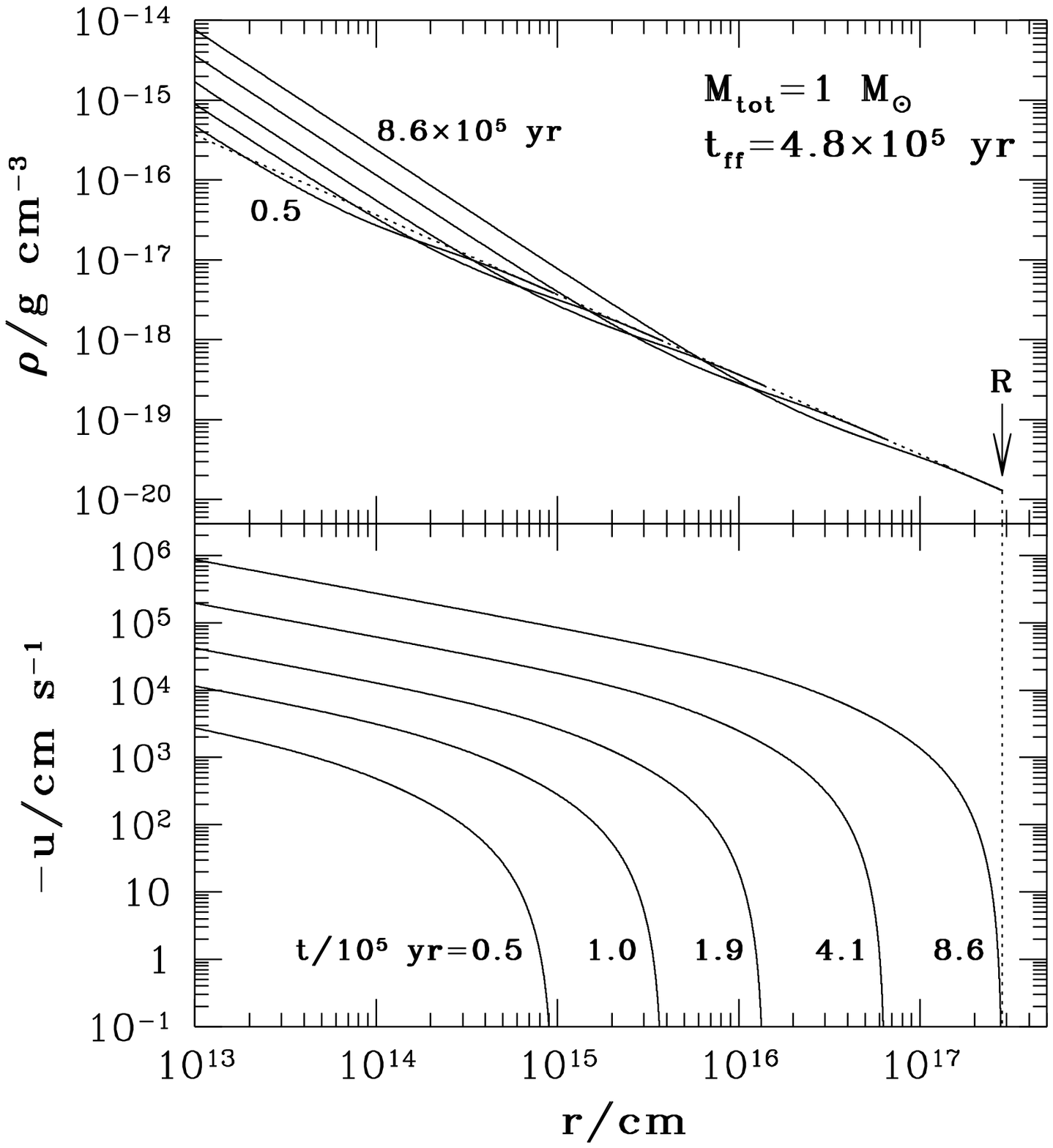]{Time evolution (from left to right), in dimensional
variables, of the density profile and velocity field for a collapsing
logotrope of $1M_\odot$ with a kinetic temperature $T=10$ K and under a
surface pressure $P_s=1.3\times10^5\ k\ {\hbox{cm}}^{-3}\ {\hbox{K}}$.  The
solid curves lie in the interior of the expansion wave (and therefore
correspond to the heavy black lines in Figs.~\ref{fig2} and \ref{fig3}), and
the broken line in the top panel traces the hydrostatic $r^{-1}$ density
profile out to the boundary $R$ of the initial cloud. The times shown are
roughly the same $t/\overline{t}_{\rm{ff}}$ used to locate $X$ in
Fig.~\ref{fig3}.
\label{fig5}}

\clearpage

\plotone{collf1.eps}

\clearpage

\plotone{collf2.eps}

\clearpage

\plotone{collf3.eps}

\clearpage

\plotone{collf4.eps}

\clearpage

\plotone{collf5.eps}

\end{document}